\documentclass{aa}
\usepackage{txfonts}
\usepackage{natbib}
\bibpunct{(}{)}{;}{a}{}{,}
\usepackage{graphicx}

\newcommand{\lL}{\ifmmode \log \frac{L}{L_{\sun}} \else $\log \frac{L}{L_{\sun}}$\fi}

\newcommand{\kms}{km~s$^{-1}$}
\newcommand{\msun}{M$_{\sun}$}
\newcommand{\zsun}{Z$_{\sun}$}
\newcommand{\lya}{Ly$\alpha$}
\newcommand{\ha}{H$\alpha$}
\newcommand{\hb}{H$\beta$}
\newcommand{\hii}{\ion{H}{ii}}
\newcommand{\heiiuv}{\ion{He}{ii}~1640}
\newcommand{\heiiopt}{\ion{He}{ii}~4686}
\newcommand{\civopt}{\ion{C}{iv}~5801-12}

\newcommand{\civuv}{\ion{C}{iv}~1550}
\newcommand{\nivuva}{\ion{N}{iv}~1486}
\newcommand{\nivuvb}{\ion{N}{iv}~1720}
\newcommand{\ovuv}{\ion{O}{v}~1371}
\newcommand{\nvuv}{\ion{N}{v}~1240}
\newcommand{\blueblue}{\ion{N}{iii}~4640+\ion{C}{iii}~4650}

\begin{document}

\title{Inferring the presence of very massive stars in local star-forming regions}
\author{F. Martins\inst{1}
\and D. Schaerer\inst{2,3}  
\and R. Marques-Chaves\inst{2}
\and A. Upadhyaya\inst{2}
}
\institute{LUPM, Universit\'e de Montpellier, CNRS, Place Eug\`ene Bataillon, F-34095 Montpellier, France  \\
Observatoire de Genève, Université de Genève, Chemin Pegasi 51, 1290, Versoix, Switzerland   \\  
CNRS, IRAP, 14 Avenue E. Belin, 31400, Toulouse, France
}

\offprints{Fabrice Martins\\ \email{fabrice.martins@umontpellier.fr}}

\date{Received / Accepted }

\abstract
{Very massive stars (VMS) have masses in excess of 100~\msun\ and are rare. However, owing to their powerful winds, very high luminosity, and efficient nucleosynthesis, they are key players of star-forming regions. In particular, their strong ionizing fluxes impact the surrounding interstellar medium.}
{We aim at detecting VMS in local star-forming region from the imprint they leave on the integrated UV and optical light.}
{We analyzed a sample of 27 star-forming regions and galaxies in the local Universe. We selected sources with a metallicity close to 12+log(O/H)$=8.3$, which is typical of the Large Magellanic Cloud. We defined empirical criteria to distinguish sources dominated by  VMS and Wolf-Rayet stars (WR), using template spectra of VMS- and WR-dominated regions. We subsequently built population synthesis models with an updated treatment of VMS. We investigated the successes and failures of these new models in accounting for the UV-optical spectroscopy of our sample sources.}
{We show that the UV range alone is not sufficient to distinguish between VMS- and WR-dominated sources because their spectra are almost identical in this range. The region of the WR bumps in the optical breaks the degeneracy. In particular, the morphology of the blue bump at 4640-4686~\AA\ is a key diagnostic.
Beyond the prototypical R136 region, which contains VMS, we identify two galaxies showing clear signatures of VMS. In two other galaxies or regions the presence of VMS can be suspected, as already discussed in the literature.
The stellar population is clearly dominated by WR stars in seven other sources. 
The most recent BPASS population synthesis models can neither account for the strong \heiiuv\ emission, nor for the shape of the blue bump in VMS- and WR-dominated sources. Our models that include VMS more realistically reproduce the UV-optical spectra of VMS-dominated sources.}
{We conclude that VMS are present in some local star-forming regions, but that separating them from WR-dominated populations requires optical spectroscopy with a high signal-to-noise ratio. A high equivalent width of \heiiuv\ is not a sufficient condition for identifying VMS. 
Populations synthesis models need to take VMS into account by incorporating not only evolutionary tracks, but also dedicated spectral libraries. Finally, we stress that the treatment of WR stars needs to be improved as well. }

\keywords{Stars: massive -- Galaxies: starburst}

\authorrunning{Martins et al.}
\titlerunning{VMS in local starbursts}

\maketitle

\section{Introduction}
\label{s_intro}

Massive stars have masses in excess of about 10~\msun\ and share specific properties. They are the main contributors to the nucleosynthesis of elements from oxygen to iron \citep[e.g.,][]{mm00,langer12}. Through their strong radiatively driven stellar winds \citep{cak,puls00}, they expel these products in their environment and actively contribute to the chemical evolution of galaxies. Massive stars also end their lives in core-collapse supernovae, which are sometimes associated with gamma-ray bursts, and they produce neutron stars or stellar black holes. When bound in binary systems, these remnants may merge and release gravitational waves \citep[e.g.,][]{abbott17}. Massive stars with initial masses higher than 100~\msun\ are a separate category. They are known as very massive stars \citep[VMS;][]{vink15}.

The most massive stars known are members of a few young massive clusters. Once thought to be a single object with a mass in excess of 1000~\msun\ \citep{feitzinger80,cassi81}, R136 at the core of the giant \hii\ region 30~Dor in the Large Magellanic Cloud (LMC) is now resolved into hundreds of massive OB stars. VMS have been observed individually \citep{mh98}. Analyses of their UV and optical spectra with non-LTE atmosphere models have established that the most massive stars of R136 exceed 100~\msun. \citet{crowther10} claimed masses between 165 and 320~\msun\ for the four most luminous objects. These values have since been revised downward \citep{besten20,brands22}, and the detection of faint companions by \citet{kalari22} establishes a maximum mass of 200~\msun\ for the most massive star. Near R136, at a projected distance of 29~pc, VFTS~682 is another VMS with an estimated mass of about 150~\msun. Astrometry indicates that it is probably a runaway star ejected from R136 \citep{banerjee12,renzo19}. The young cluster NGC~3603 in the Milky Way is also dominated by a few VMS \citep{cd98,crowther10}. Its most massive components reach 150~\msun, and one of them is a binary star \citep{schnurr08}. The Arches cluster near the Galactic center hosts a rich population of massive OB stars, some of which reach masses of 120~\msun\ \citep{arches}. In the Local Group, M33 hosts the giant \hii\ region NGC~604, in which several candidate VMS have been identified by \citet{bruw03}.

Very massive stars are characterized by spectra that are dominated by wind features. All of them are classified as WNh objects, that is, Wolf-Rayet (WR) stars of the WN sequence, showing signs of hydrogen at their surface \citep{ssm96}. WR stars are the descendants of OB stars and are chemically evolved, showing signs of hydrogen and helium burning. However, despite their spectral classification, VMS are not genuine WR stars: most of them are main-sequence objects \citep{arches,besten20,brands22}. VMS only look like WR stars because of their strong stellar winds, which cause most lines to appear as either P-Cygni or emission lines. VMS have higher mass-loss rates than their lower-mass OB counterparts, but the mass loss is comparable to that of WR stars \citep{arches,gh08}. The proximity to the Eddington limit favors multiple scattering of photons and thus more efficient momentum transfer, leading to higher mass-loss rates \citep{graef11,vink11,sander20b,besten20massloss,graef21}.

The stronger stellar winds of VMS alter their evolution, as was shown by \citet{graef21}. The winds also lead to specific spectral features \citep{gv15}. \citet{mp22} showed that consistent evolutionary and atmosphere calculations for VMS lead to synthetic UV and optical spectra that are comparable to those of known VMS. The authors identified \heiiuv, \nivuva,\ and \nivuvb\ as features that are visible in very massive main-sequence objects, but not in normal OB stars. These features are also seen in normal WR stars, but at later stages in the stellar evolution. VMS typically live 2 to 3~Myr \citep{yusof13,kohler15,chen15,mp22}, while WR stars produced by single stars appear after $\sim$3~Myr and are present up to 5-6~Myr in starbursts (up to 8~Myr when binarity is involved), depending on metallicity \citep{meynet03,meynet05,sv98,bpass}. The equivalent width (EW) of \heiiuv\ in VMS can reach 30~\AA\ \citep{mp22}. VMS can also display \heiiopt\ and \civopt\ emission. \citet{crowther16} built the integrated UV spectrum of R136 from individual spectra. They showed that the UV flux is dominated by the seven most massive stars. In particular, \heiiuv\ is only produced by VMS. \citet{mp22} confirmed that VMS can drastically modify the morphology of the UV (and optical) spectra of young starbursts. 

Very massive stars have thus caught attention because some of the broad and intense emission in star-forming clusters might be explained when they are included in a model. In particular, \heiiuv\ is sometimes stronger than any population synthesis can predict \citep{leitherer20}. Population synthesis models usually incorporate stars only up to 100~\msun\ and thus exclude VMS. The similarity between the observed integrated spectrum of clusters NGC~3125-A1, NGC~5253-5, and II~Zw~40-N on one hand and the VMS in R136 on the other hand caused \citet{wofford14,wofford23}, \citet{smith16}, and \citet{leitherer18} to speculate that these clusters might host unresolved VMS. \citet{mp22} showed that including VMS in population synthesis models improved the fit of the UV spectrum of NGC~3125-A1. \citet{senchyna20} argued that only VMS could explain the strong EW(\heiiuv) in their sample of starburst galaxies. To reach this conclusion, they relied on an updated version of the population synthesis code of \citet{bc03}, described in \citet{vg17}, and they focused on the EW of \heiiuv.  The current set of BPASS models \citep{bpass} incorporates a partial treatment of VMS: it includes evolutionary tracks for masses above 100~\msun, but the synthetic spectra for this mass range are the same as those of lower mass stars. We show below that this affects the appearance of starburst regions. Finally, using updated population synthesis models (see Sect.~\ref{s_popsyn}), \citet{mestric23} showed that the intense \heiiuv\ and \ion{N}{iv} emission lines of the lensed starburst cluster Sunburst at z=2.37 \citep{dahle16,vanzella22} can only be explained if VMS are included in the modeling. Although less numerous than expected from a simple extrapolation of the IMF, the VMS in Sunburst contribute an additional 15\% of the Lyman-continuum photons, which boosts the ionizing power of the population. This stresses the key role that VMS may have played in the early Universe, where young stellar population are being uncovered by the JWST close to the reionization epoch \citep{castellano22,bunker23,atek23}.

We investigate the presence of VMS in local star-forming regions. 
Our primary goal is to identify star-forming regions 
or galaxies in which VMS are present and dominate the integrated UV and optical light. This also implies separating sources dominated by VMS from those dominated by classical WR stars, as we show below. We stress that our approach likely misses VMS in regions in which they are mixed with other populations and in which their number is not sufficient for them to affect the integrated light. Our investigation of VMS in star-forming regions therefore leads to conservative results. 
We describe our sample selection in Sect.~\ref{s_2}. We build an empirical classification scheme to infer the presence of VMS and distinguish them from WR stars in  Sect.~\ref{s_emp}. We build population synthesis models including VMS in Sect.~\ref{s_popsyn}, where we discuss the shortcomings of current models.  We discuss our results in Sect.~\ref{s_disc}. We give our conclusions in Sect.~\ref{s_conc}.

\section{Sample and observational data}
\label{s_2}

The best template spectrum for a region hosting VMS sources is R136 in the LMC (see Sect.~\ref{s_emp}). In addition, \citet{mp22} calculated VMS models that were tuned to a metallicity of 1/2.5~\zsun. Because we rely on both sets of spectra (template + synthetic) to identify VMS-hosts, we selected sources with metallicities 12+log(O/H) = 8.29$\pm$0.1, which is appropriate for the LMC. Our primary criterion to build the sample was the availability of both UV and optical spectroscopic data because the two spectral regions are necessary to firmly establish the presence of VMS, as we show below. We compiled star-forming regions in the local Universe with a metallicity similar to that of the LMC.

 Our sample is a mixture of various star-forming regions, from clusters to star-forming knots within galaxies and entire starburst galaxies. We stress that by construction, this sample is not homogeneous in terms of physical properties, except for the metal content and their nature (star-forming regions). The sample includes the starburst clusters NGC~3125-A1, NGC~5253-5, the giant star-forming region Tol89-A, and six \ion{H}{ii} regions in M101 (NGC~5447, NGC~5455, NGC~5462, SIP2007-1, HGGK~1054, and HGGK~1216). Three additional star-forming complexes in NGC~4214, NGC~4670 and Mrk~33 were included. The rest of the sample includes sources from the CLASSY HST legacy survey\footnote{\url{https://www.danielleaberg.com/classy}\newline \url{https://archive.stsci.edu/hlsp/classy}} as well as five sources from \citet{senchyna17,senchyna20}. These sources are usually star-forming knots in local starburst galaxies, that is, the size of the star-forming region is smaller than that of the host galaxy. For two sources, J1200 and J1428, the spectra are those of the entire galaxy. 
Our final sample is made of 27 sources. The full list together with the measured metallicities (obtained from the direct method), distances, and references is given in Table~\ref{tab_1}.
 
For the sources studied by Senchyna et al.,\ we added their SB number in their ID (i.e., J0115$-$0051/SB49 for source SB49 of Senchyna et al.). In the following, we refer to each source with SDSS IDs by the first four digits of its ID (i.e., J1105 for J1105$+$4444). The metallicities of SIP2007~1 and HGGK~1054 were estimated from the position in M101 and the metallicity gradient determined by \citet{esteban20}. For the sources of M101 in NGC~5447 and NGC~5462, we added an ``S'' identifier to specify that the source is the most southern knot in the star-forming complexes of these regions. 

The UV and optical data available for the sample sources come from various telescopes and instruments. They were retrieved from science archives or by direct contact with the principal investigators of the observations. The list of observations is given in Table~\ref{tab_2}, which also includes the spectral resolution of the data and references to publications making use of them. \\
The UV HST data were downloaded from the MAST archive, and we used them without further processing, that is, we use the reduced data available in the archives. Some of the ESO data are science-ready data available from the phase 3 archive, and in those cases, we used them without further processing. The FORS data for NGC~3125-A1 are raw data that were reprocessed with the public pipeline in the esoreflex environment \citep{reflex}. Standard reduction was performed, and we extracted the optical spectrum at the position of cluster A1 over a  region of 10 pixels, corresponding to 1.25\arcsec. Both the UV and optical spectra encompass cluster A1 and A2 \citep[see Fig.2 of][]{wofford23}. A1 dominates the integrated UV light, but A2 contributes about one-third$^{}$ of the optical flux near 7200~\AA\ \citep{sidoli06}.
The UVES data of NGC5253-5 were processed with the ESO pipeline in the reflex environment. Two-dimensional data were created, and the spectrum of cluster 5 was extracted from these data (in a spatial region of 6 pixels, corresponding to 0.96\arcsec).
A MUSE data cube is available in the archive for J1304. We extracted the spectrum over a 5-pixel region centered on the target, corresponding to 1\arcsec. Similarly, we extracted the optical spectrum of R136 from the available MUSE data cube using a 30-pixel aperture (see below).

All spectroscopic data were manually normalized to the continuum by selecting regions as free as possible of stellar absorption or emission lines. In the following, we show that this process was not obvious when signal-to-noise ratio (SNR) was low. This affected the measurements of equivalent widths (EW).

\begin{figure}[t]
\centering
\includegraphics[width=0.49\textwidth]{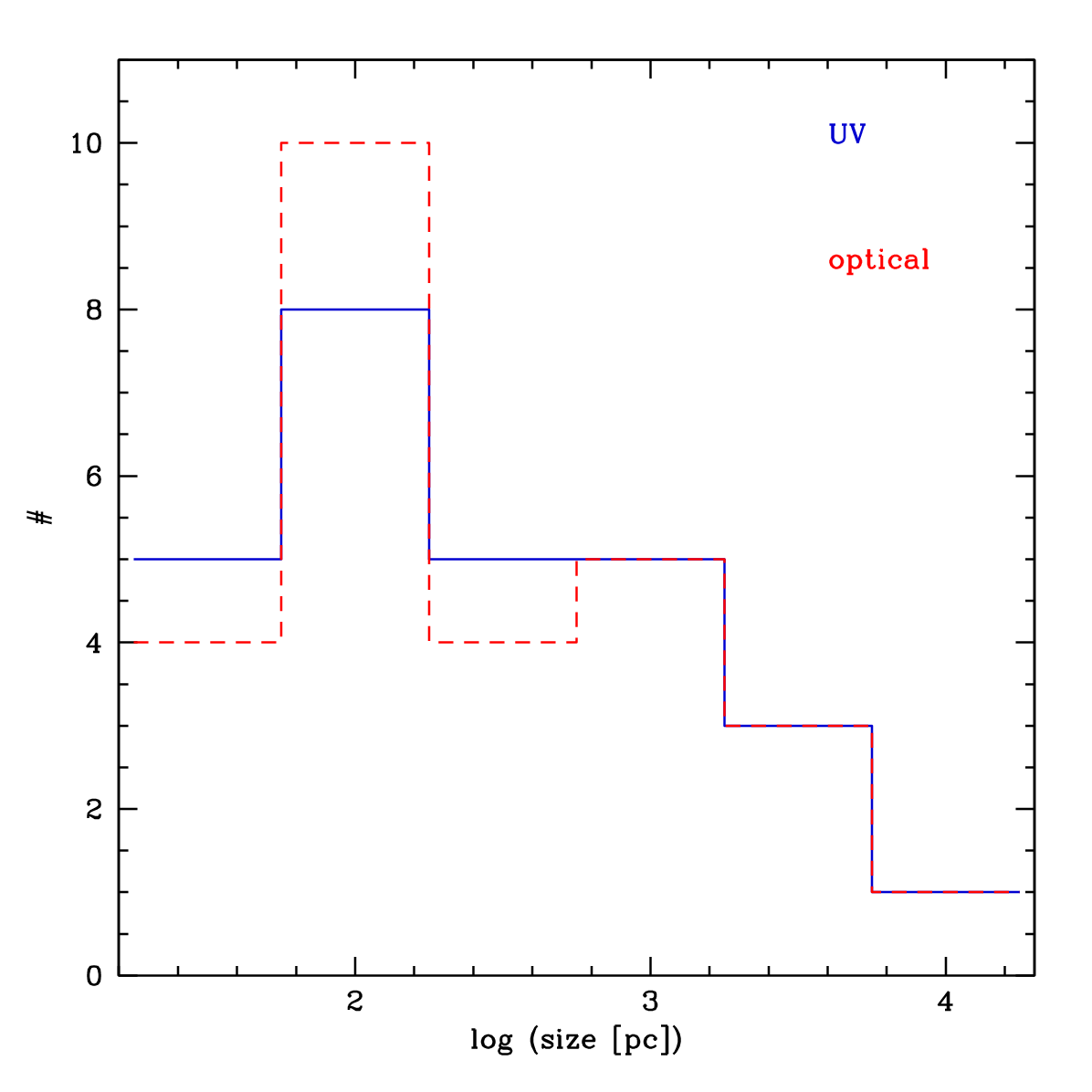}
\caption{Number of sources as a function of the physical size covered by their UV (solid blue line) and optical (dashed red line) observational data.}
\label{fig_hist_size}
\end{figure}

\begin{table*}[ht]
\begin{center}
\caption{Sources, position, metallicity, and reference for metallicity.} \label{tab_1}
\begin{tabular}{lcccccc}
\hline
Source                 &   RA          & DEC      & 12+log(O/H) & reference  & distance & reference  \\
                       &               &          &             &   metallicity         & [Mpc]    &    distance        \\
\hline
R136                    & 05 38 42.40  &  -69 06 03.36  & 8.40  &  (1)       &  0.0496      & (13) \\
NGC~3125-A1             & 10 06 33.70  &  -29 56 10.40  & 8.32  &  (2)       &  11.5-14.84  & (14,15) \\
NGC~5253-5              & 13 39 55.63  &  -31 38 28.34  & 8.26  &  (3)       &  3.55        & (16) \\
Tol~89-A                & 14 01 20.09  &  -33 04 10.70  & 8.27 & (4)         &  14.7        & (14)\\
NGC~5447S               & 14 02 30.60  &   54 16 09.72  & 8.30  &  (5)       &  6.7         & (17) \\
NGC~5455                & 14 03 01.16  &   54 14 29.16  & 8.30-8.42  &  (5)  &  6.7         & (17) \\
NGC~5462S               & 14 03 53.13  &   54 22 06.52  & 8.25  &  (5)       &  6.7         & (17)\\
SIP2007-1              & 14 03 31.30  &   54 21 14.80  & $\sim$8.5   & (5)   &  6.7         & (17)   \\
HGGK~1054              & 14 03 34.00  &   54 18 37.06  & $\sim$8.4   & (5)   &  6.7         & (17)   \\
HGGK~1216              & 14 04 11.21  &   54 25 18.16  & 8.18   &  (5)       &  6.7         & (17) \\
NGC~4214                & 12 15 39.48  &   36 19 35.36  & 8.30   &  (6)      &  2.95        & (18) \\
NGC~4670                & 12 45 17.27  &   27 07 32.13  & 8.17-8.27 & (7)    &  19.6        & (19) \\
Mrk~33                  & 10 32 31.81  &   54 24 03.51  & $\sim$8.3 & (8)     &  15.4        & (20) \\
J0036$-$3333           & 00 36 52.68   &  -33 33 17.24  & 8.21   & (9)       &  89          & (9)\\
J0115$-$0051/SB49      & 01 15 33.82   &  -00 51 31.20  & 8.20   & (10)      &  31          & (10) \\
J0823$+$2806           & 08 23 54.96   &   28 06 21.60  & 8.28   & (9)       &  209         & (9) \\
J0942$+$0928/SB80      & 09 42 56.74   &   09 28 16.20  & 8.24   & (12)      & 46           & (12) \\
J1105$+$4444           & 11 05 08.16   &   44 44 47.40  & 8.23   & (9)       & 93           & (9) \\
J1129$+$2034/SB179     & 11 29 14.15   &   20 34 52.01  & 8.28-8.30   & (9,10) & 21-25      & (9,12)  \\
J1132$+$1411/SB125     & 11 32 35.35   &   14 11 29.83  & 8.25   & (9)       & 76-78        & (9,10) \\
J1157$+$3220           & 11 57 31.68   &   32 20 30.12  & 8.43   & (9)       & 48           & (9) \\
J1200$+$1343           & 12 00 33.42   &   13 43 07.95  & 8.26   & (9)       & 300          & (9) \\
J1215$+$2038/SB191     & 12 15 18.60   &   20 38 26.70  & 8.30 & (12)        & 10           & (12) \\
J1304$-$0333/SB9       & 13 04 32.27   &  -03 33 22.10  & 8.30   & (10)      & 20           & (10)\\
J1314$+$3452/SB153     & 13 14 47.37   &   34 52 59.81  & 8.26   & (9)       & 13-11        & (9,10) \\
J1428$+$1653           & 14 28 56.41   &   16 53 39.32  & 8.33   & (9)       & 880          & (9) \\
J1525$+$0757           & 15 25 21.84   &   07 57 20.30  & 8.33   & (9)       & 343          & (9) \\
\hline
\end{tabular}
\tablefoot{Columns are the source ID, the right ascension, the declination, the metallicity, the reference for the metallicity, the distance, and the reference for the distance.
  1- \citet{dopita19}; 2- \citet{hc06}; 3- \citet{monreal12}; 4- \citet{sidoli06}; 5- \citet{esteban20}; 6- \citet{ks96}; 7- \citet{kumari18}; 8- \citet{me00}; 9- \citet{berg22}; 10- \citet{senchyna20}; 11- \citet{james09}; 12- \citet{senchyna17}; 13- \citet{piet19}; 14- \citet{schaerer99b}; 15- \citet{mould00}; 16- \citet{tully13}; 17- \citet{freedman01}; 18- \citet{jacobs09}; 19- \citet{bottinelli84}; 20- \citet{tully16}.}
\end{center}
\end{table*}

\begin{table*}[ht]
\begin{center}
\caption{Observational data} \label{tab_2}
\begin{tabular}{l|cccc|ccccc}
\hline
Target         &       & UV       &  & & & optical &   \\
\hline
                       &  source  & ID & reference & resolution & source & ID & reference & resolution \\
                &      &          &         & \AA & & & & \AA \\
\hline
      R136             &  HST/STIS     &  12465   & (1)        & 1.2 & ESO/MUSE & 0104.D-0084    & (2)   & 2.3\\
NGC~3125-A1             &  HST/COS      &  15828    & (4)      &  0.08 & ESO/FORS & 074.B-0108           &  (5) &  15 \\
NGC~5253-5              &  HST/STIS    &  8232     & (6)       &  1.8 & ESO/UVES      & 073.B-0283   & (6)  & 0.2  \\
Tol~89-A                &  HST/STIS    &  7513     & (7)       &  3.1 & ESO/UVES      & 073.B-0238   & (7)  & 0.2  \\
NGC~5447S               &  HST/COS     &  15126    &           &  0.6 & GTC/OSIRIS & GTC14-16A & (8)  & 6.52  \\
NGC~5455                &  HST/COS     &  15126    &           &  0.6 & GTC/OSIRIS & GTC14-16A  & (8) & 6.52 (B) and 2.46 (V)   \\
NGC~5462S               &  HST/COS     &  15126    &           &  0.6 & GTC/OSIRIS & GTC14-16A & (8)  & 6.52 (B) and 2.46 (V)  \\
SIP2007-1              &  HST/COS     &  15126    &           &  0.6  & SDSS   &      & (9)  & 2.5   \\
HGGK~1054              &  HST/COS     &  15126    &           &  0.6 & SDSS  &       & (9)  & 2.5   \\
HGGK~1216              &  HST/COS     &  15126    &           &  0.6 & GTC/OSIRIS & GTC14-16A   & (8)   & 6.52 (B) and 2.46 (V) \\
NGC~4214                &  HST/STIS &  9036   & (10,11)    &   1.2 & HST/STIS & 15846   & &   0.56     \\
NGC~4670                &  HST/COS     &  15193     & (12)    &  0.08 & HST/STIS & 15846  &  &  0.56      \\
Mrk~33                  &  HST/STIS    &  9036     & (10,11)   &  1.2 & SDSS   &     & (9)  & 2.5  \\
J0036$-$3333           &  HST/COS     &  13017    & (13,14,15)&   0.073 & SDSS  &      & (9)  & 2.5   \\
J0115$-$0051/SB49              &  HST/COS     &  15185  & (16)&  0.073  & SDSS   &      & (9) & 2.5    \\
J0823$+$2806           &  HST/COS     &  13017    & (12,13,14) &  0.073 & SDSS   &     & (9)  & 2.5   \\
J0942$+$0928/SB80             &  HST/COS     &  14168    & (17)       & 0.073 & SDSS   &     & (9) & 2.5    \\
J1105$+$4444           &  HST/COS     &  15840 & (14)&  0.073 & SDSS  &     & (9) & 2.5    \\
J1129$+$2034/SB179         &  HST/COS     &  15840    & (13,14)    & 0.2 & SDSS  &      & (9) & 2.5    \\
J1132$+$1411/SB125     &  HST/COS     &  15840    & (13,14) & 0.073 & SDSS   &     & (9)  & 2.5   \\
J1157$+$3220           &  HST/COS     &  15840    & (13,14)& 0.073 & SDSS   &    & (9)  & 2.5   \\
J1200$+$1343           &  HST/COS     &  15840    & (13,14)   & 0.073  & SDSS   &     & (9) & 2.5    \\
J1215$+$2038/SB191            &  HST/COS     &  14168    & (17)      & 0.073  & SDSS    &    & (9) & 2.5    \\
J1304$-$0333/SB9              &  HST/COS     &  15185    & (16)      & 0.073 & ESO/MUSE    &  0104.D-0199  &  & 2.3  \\
J1314$+$3452/SB153      &  HST/COS     &  15840  & (13,14)   & 0.073 & SDSS    &    & (9)  & 2.5   \\
J1428$+$1653           &  HST/COS     &  13017    & (13,14,15) &  0.2 & SDSS   &     & (9)  & 2.5   \\
J1525$+$0757           &  HST/COS     &  13017    & (13,14,15) &  0.2 & SDSS   &     & (9)  & 2.5   \\
\hline
\end{tabular}
\tablefoot{Columns are the source ID, the source of UV data, the program ID and reference, the UV spectral resolution, the source of the optical data, the program ID and reference, and the optical spectral resolution. For OSIRIS data, the optical spectra are separated into B and V band and have a different resolution.
1- \citet{crowther16}; 2- \citet{castro18}; 3- \citet{chandar04}; 4- \citet{wofford23}; 5- \citet{hc06}; 6- \citet{smith16}; 7- \citet{sidoli06}; 8- \citet{esteban20}; 9- \citet{sdss}; 10- \citet{chandar03}; 11- \citet{chandar04}, 12- \citet{hernandez20} 13- \citet{berg22}; 14- \citet{james22}; 15- \citet{heckman15}; 16- \citet{senchyna20}; 17- \citet{senchyna17}. }
\end{center}
\end{table*}

Table~\ref{tab_2} shows that some data are taken through slits, some through fibers, and some through integral field units. The geometry of the area that is covered by the observations is thus not homogeneous. Table~\ref{tab_3} provides information on the spatial extent of the observations in physical units, using the distances listed in Table~\ref{tab_1} and the observational configuration. The information is given for both UV and optical observations\footnote{For STIS data, the sizes were retrieved from the EXTRSIZE (size of extraction box) and PLATESC keywords of the headers.}.  Fig.~\ref{fig_hist_size} shows that our sample is mostly made of sources with sizes smaller than 1.5~kpc, with a peak near 50-150~pc. This obviously means that the sample includes a variety of star-forming complexes, from single clusters to knots that likely contain several star-forming entities. More information on the morphology of the sample sources is given in Appendix~\ref{ap_spatial}. For the sources that are part of the CLASSY and Senchyna et al. samples, we refer to the original publications, where images of the regions covered by the observations can be found. We stress that the observations focused on specific regions of the host galaxies even in these objects. Only for J1200 and J1428 is the entire galactic light encompassed in the observational entrance feature.

\begin{table*}[ht]
\begin{center}
\caption{Physical size and estimated mass of the sample sources.} \label{tab_3}
\begin{tabular}{lcccccc}
\hline
source              &   size UV & size optical  &  mass  & reference\\
                    &   [pc]    &    [pc]       &  [\msun]  & for mass  \\
\hline
R136                &   0.5     & 0.5          &  $\leq 10^{4.70}$ & (1)\\
NGC~3125-A1         &   139 and 180 & 70x45 and 90x58  &  $10^{5.62}$ & (2) \\        
NGC~5253-5          &   9.3x1.7  & 56x24       &  $<10^{5.52}$ & (3)  \\
Tol~89-A            &   20x35    & 171x100     & $10^{5.08}-10^{-5.48}$ & (4) \\  
NGC~5447S           &    81       & 198x26     & -- & -- \\          
NGC~5455            &     81       & 165x26    & -- & --  \\           
NGC~5462S           &    81       & 133x26     & -- & -- \\          
SIP2007-1           &    81       & 97     & -- & -- \\          
HGGK~1054           &    81       & 97      & -- & -- \\          
HGGK~1216           &    81       & 117x26     & -- & -- \\          
NGC~4214            &   5x3         &   7x3    & -- & -- \\          
NGC~4670            &    238       &  47x19    & -- & -- \\                       
Mrk~33                   & 20x15       & 224    & --  & -- \\             
J0036$-$3333            & 1078        &   1294     &  $10^{8.77}$ & (5, COS) \\       
J0115$-$0051/SB49       & 376         & 451     &   $10^{5.80}$ & (6, SDSS) \\         
J0823$+$2806            &  2530       & 3040    & $10^{8.47}$ & (5, COS)  \\ 
J0942$+$0928/SB80       &  557        &  669    & $10^{6.10}$ & (7, SDSS) \\  
J1105$+$4444            & 1127        & 1352 &  $10^{7.76}$ & (5, COS) \\
J1129$+$2034/SB179      & 254 and 303     &   305 and 364 &  $10^{5.75}$ and  $10^{5.20}$ & (5, COS) and  (7, SDSS)  \\  
J1132$+$1411/SB125      & 921 and 945     & 1105 and 1134  &  $10^{6.79}$ and $10^{6.66}$ & (5, COS) and (6, SDSS) \\        
J1157$+$3220            & 582         &   698    &  $10^{7.33}$ & (5, COS) \\  
J1200$+$1343            &  3636       &  4363    &  $10^{7.46}$ & (5, COS) \\  
J1215$+$2038/SB191      &   121       &  145     & $10^{4.9}$ & (7, SDSS)\\     
J1304$-$0333/SB9        &    242      &   97  & $10^{5.05}$ & (6, SDSS) \\  
J1314$+$3452/SB153      &   158 and 133   & 189 and 160  &  $10^{5.28}$ and $10^{4.93}$ & (5, COS) and (6, SDSS) \\ 
J1428$+$1653            &  10666      &   12800  &  $10^{9.50}$ & (5, COS)  \\        
J1525$+$0757            &   4157      &   4989   &  $10^{9.13}$ & (5, COS)  \\   
\hline
\end{tabular}
\tablefoot{The sizes are those of the region over which the spectrum is extracted.  For SDSS and COS data, we give the diameter (3\arcsec\ and 2.5\arcsec) at the distance of the source. For slit spectra, we give the extraction size times the slit width at the distance of the source. Multiple entries separated by a slash indicate multiple values of the source distance (see Tab.~\ref{tab_1}). Columns 2 and 3 correspond to UV and optical data, respectively. The masses in column 4 are taken from the literature. The labels COS or SDSS for masses refer to masses extracted from an SED fitting of photometry in the COS or SDSS aperture. Masses without a label are for the entire star-forming region. The references are 1- \citet{crowther16}; 2- \citet{hc06}; 3- \citet{smith16}; 4- \citet{sidoli06}; 5- \citet{berg22}; 6- \citet{senchyna20}; 7- \citet{senchyna17}.  }
\end{center}
\end{table*}

\section{Empirical study and classification}
\label{s_emp}

In this section, our goal is to use template spectra of star-forming regions whose stellar content is known to infer VMS in the integrated light of unresolved star-forming regions.
R136 in the LMC serves as a template for star-forming regions hosting VMS. Its massive stellar content is known (see \citealt{crowther16,besten20,brands22} for recent studies): VMS are present, and given the age of the cluster (1 to 2~Myr), no classical WR stars are detected. The integrated UV spectrum of R136 was presented by \citet{crowther16}. For the optical spectrum, MUSE observations of the cluster by \citet{castro18} are available in the ESO archive (see Table~\ref{tab_2}). We extracted the spectrum using a circular region of 30 pixels centered on R136 to cover the same region as the UV spectrum. The result is shown in Fig.~\ref{fig_emp}. We stress that R136 is much smaller by far than any of the sources of our sample, and it is therefore not representative of their spatial morphology and potential multicomponent nature (see Table~\ref{tab_3}). However, this integrated spectrum is dominated by OB and VMS stars and thus avoids contributions from WR stars, which are found on larger scales in the 30 Dor region.
Regardless of its spatial morphology, any source that displays a UV+optical spectrum similar to that of R136 is thus very likely dominated by VMS, whether another type of object is present or not. The presence of VMS is thus inferred from the spectral resemblance to R136.

Similarly, a template for star-forming regions that host only WR (and OB) stars but no VMS would be desirable, and for which integrated UV and optical spectra are available.
Unfortunately, an equivalent of R136 like this does not exist for WR stars. 

The giant \ion{H}{ii} region NGC~604 in M33 hosts a population of massive stars that includes WR stars of the WN and WC subtypes \citep{cm81,dod81,drissen90,drissen93}. Most stars have been resolved spectroscopically \citep{drissen08}, and some of them have been placed in an HR diagram \citep[e.g.,][]{eldridge11}. \citet{terlevich96} and \citet{gonzalez00} presented integrated UV and optical spectra covering most of the massive-star population. NGC~604 thus appears as an attractive region to serve as a template for a starburst dominated by WR stars. However, red supergiants are also present in the region \citep{terlevich96}. Finally, \citet{bruw03} reported stars with masses in excess of 100~\msun. HST STIS spectra of these objects revealed \heiiuv\ emission, as seen in VMS. Consequently, NGC~604 hosts a mixture of WR, VMS, and red supergiants, likely indicating an extended episode of star formation. NGC~604 is thus a good template for studies of unresolved star-forming regions, but it is not suited for our purpose: both WR and VMS are present and contribute to the integrated light, so that NGC~604 cannot serve as a template for a WR-dominated population.

We therefore relied on templates of other star-forming regions in which WR stars are suspected, but not directly probed. Sources dominated by WR features are common among star-forming galaxies. \citet{brinch08} compiled hundreds of them in the SDSS survey\footnote{\url{http://www.sdss.org}}. 
At the metallicities of interest here,
the giant \ion{H}{ii} region Tol89 in the galaxy NGC~5398 stands out. It hosts several young massive clusters. Region A in these clusters shows the optical bumps that are attributed to WR stars \citep{durret85,schaerer99}. The region is made of four subclusters, A1 to A4. \citet{sidoli06} showed that WR features are present in all clusters except for A3, which is more reddened than the other sources. The authors showed that clusters A1 and A4 mainly host WN stars that produce a strong blue bump and no red bump\footnote{The blue bump is caused by \ion{N}{iii}~4640, \ion{C}{iii}~4650, and \heiiopt, while the red bump is due to \civopt.}, while cluster A2 contains both WN and WC stars, the latter driving the observed red bump. Using template spectra of individual WN5-6 and WC4 stars and adjusting the number of WN and WC stars, Sidoli et al. built an empirical integrated spectrum of the region that was able to reproduce the morphology of the optical WR bumps in Tol89-A (see their Figs.~13 and 14). Both UV and optical data are publicly available for this source, and therefore, we used Tol89-A as a reference for a population dominated by WR stars. The presence of both WN and WC stars, the two main classes of WR stars, make it a good reference template that shows all features that WR stars of different types might exhibit. In the following, we show that Tol89-A and R136 have different spectral morphologies that can be used to probe the presence of VMS and WR stars (see also Sect.~\ref{s_popshort}).

\subsection{Classification scheme}
\label{s_classif}

\begin{figure}[t]
\centering
\includegraphics[width=0.49\textwidth]{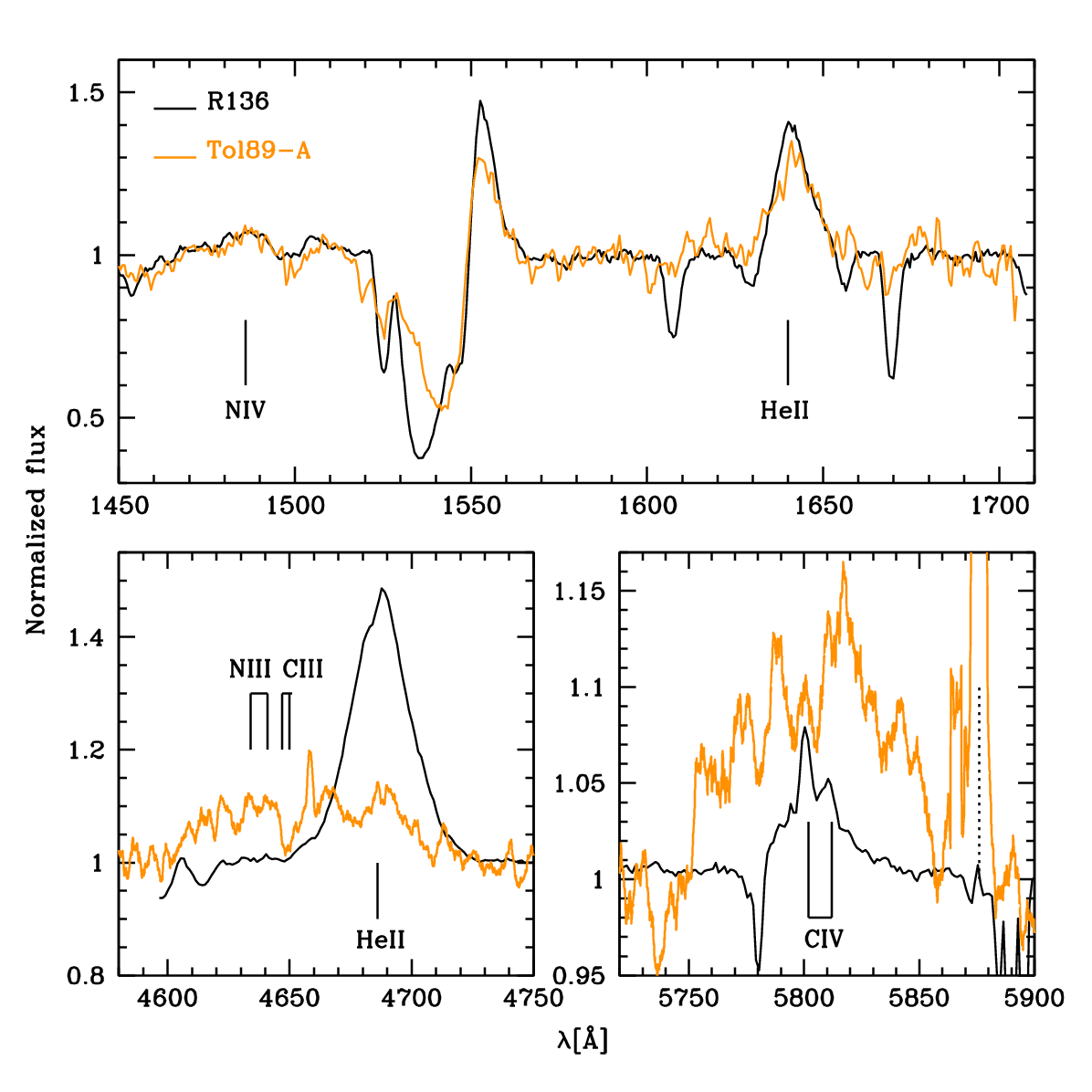}
\caption{Comparison of the observed spectrum of R136 (black) and Tol89-A (orange). The ions causing the various emission features are marked by solid lines. The dotted line indicates nebular \ion{He}{i}~5876 emission.}
\label{fig_emp}
\end{figure}

We now define the observational criteria based on which we distinguished between sources dominated by VMS or WR stars. To do this, we compare in Fig.~\ref{fig_emp} the UV spectrum and optical WR bumps of R136 and Tol89-A, our reference sources for the two cases.

In the UV range, the spectra of the two sources are remarkably similar. In particular, the \heiiuv\ emission of R136 is very similar to that of Tol89-A: the equivalent width (EW) of \heiiuv\ is larger than $\sim$3~\AA\ (taking the uncertainty into account) in both sources (see Sect.~\ref{s_ew} and Table~\ref{tab_ew2}). The line is broad in both sources. \citet{saxena20} defined a full width at half maximum  of 1000~\kms\ to separate between stellar and nebular \heiiuv: with a width of about 1500~\kms\ , R136 and Tol~89-A both fall in the first category. R136 and Tol89-A also have the same \nivuva\ morphology. 
\nivuvb\ is not covered by the available observations of these sources and was therefore not used in our classification.
We conclude that the UV range alone is not sufficient to distinguish between sources dominated by VMS or WR stars. 

In the optical range, clear differences appear in the blue bump morphology. R136 displays a strong \heiiopt\ emission and has basically no \blueblue\ emission. In contrast, Tol89-A has an \blueblue\ feature that is as intense as \heiiopt. The blue bump is thus a key feature for distinguishing sources dominated by VMS from those in which mostly normal WR stars are present. We stress that we considered relative line ratios and not absolute line strengths. \citet{leitherer19} showed that the ratio of \heiiuv\ to \heiiopt\ is rather constant among WR stars at about 8. Because \heiiuv\ is very similar in R136 and Tol89-A, the same might be expected for \heiiopt. Fig.~\ref{fig_emp} shows that this is not the case: \heiiopt\ is weaker in Tol89-A. This is likely due to the underlying stellar populations and/or nebular continuum emission (see below). However, it is clear that \blueblue\ is as strong as \heiiopt\ in Tol89-A but is much weaker in R136. This specific qualitative criterion is used below.

The red bump is caused by \civopt. In R136, the two components are narrow and resolved in moderate-resolution spectra. In Tol89-A, the two components of the doublet are broadened in the dense stellar atmosphere and merge into a single broad feature. A broad emission is thus characteristic of WR stars, while a relatively weak doublet indicates VMS. 

The spectroscopic characteristics of VMS hosts just described are also observed in individual VMS at LMC metallicity. \citet{hainich14} studied the properties of WN stars in the LMC. Apart from the stars in R136, their most luminous and most massive stars (e.g., BAT99-77 and BAT99-116) do show a weak \blueblue\ component with a relatively strong \heiiopt. Because the intensity  of the \blueblue\ emission is partly rooted in the initial amount of carbon and nitrogen, it is weaker in the LMC than in the Galaxy. \citet{ssm96} showed that the ratio of \heiiopt\ to \blueblue\ in WN stars increased with galactocentric radius in the Galaxy, and thus increased at lower metallicity (see their Fig.~11). \citet{crowther23} reported line luminosities at 4630~\AA\ (including \ion{N}{v}~4603-20 and \ion{N}{iii}~4634-41) and in \heiiopt\ for Galactic and LMC stars. The 4630/4686 luminosity ratio decreases from 0.58$\pm$0.32 in the Galaxy to 0.07$\pm$0.03 in the LMC for WN5-7h stars, which is the typical spectral type of VMS. The morphology of the blue bump of VMS in R136, that is, a strong \heiiopt\ and weak \blueblue\ emission, is thus representative of VMS at LMC metallicity.

\begin{figure}[t]
\centering
\includegraphics[width=0.49\textwidth]{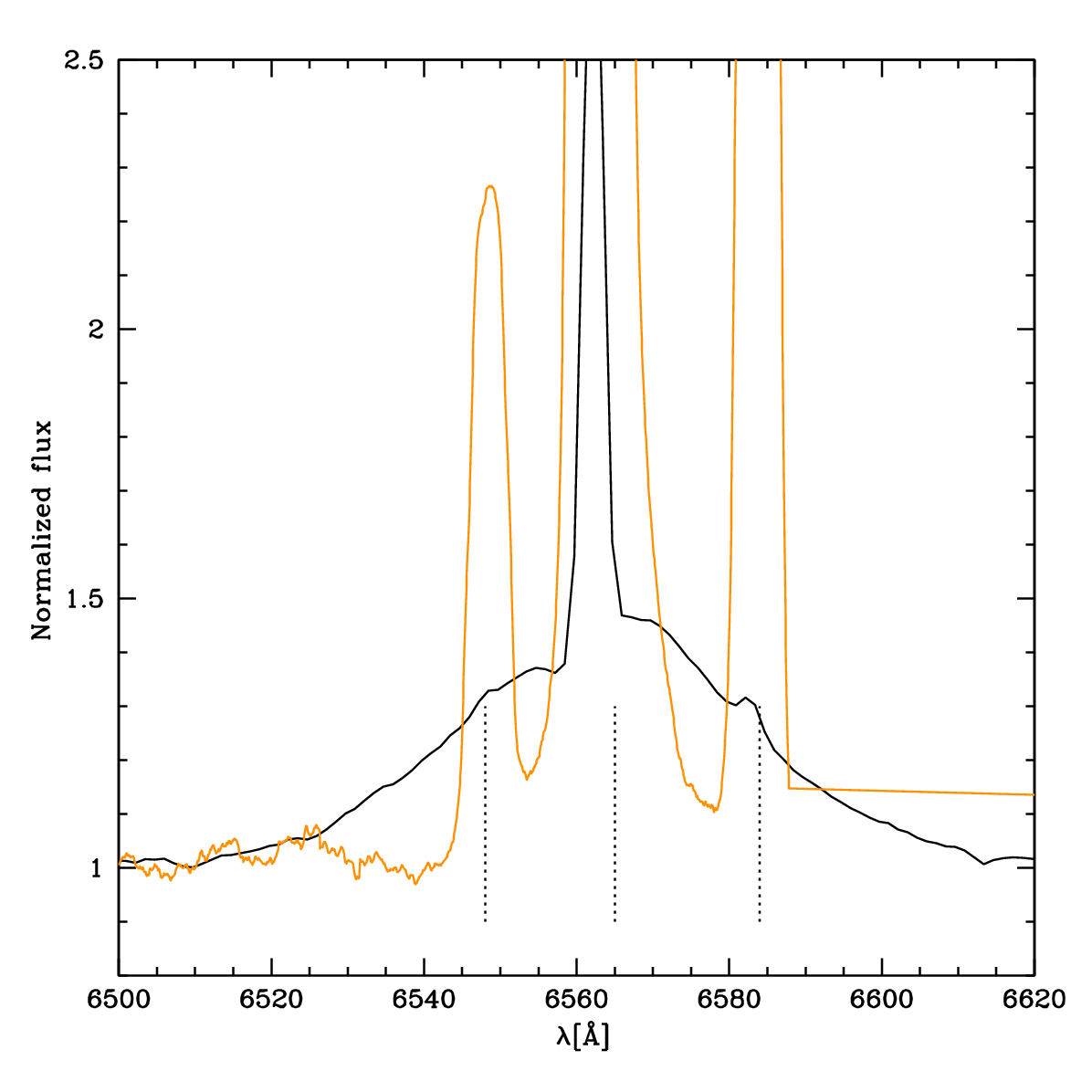}
\caption{Same as Fig.~\ref{fig_emp} for the \ha\ line. Nebular emission is indicated by vertical dotted lines.}
\label{fig_empHa}
\end{figure}

Very massive stars are spectroscopically classified as WNh stars because they contain a significant fraction of hydrogen in their atmosphere \citep{crowther10,hainich14,besten20}. In R136, the analysis of \citet{besten20} indicates a hydrogen mass fraction of about 50\% for the three most massive stars. This translates into the presence of a broad stellar \ha\ emission, as shown in Fig.~\ref{fig_empHa}. This line could thus be a diagnostic of VMS. However, even if classical WN stars have less or no hydrogen in their atmosphere, the analysis of \citet{hainich14} indicates that a significant number of LMC WR stars still contains hydrogen. The presence of \ha\ in classical WR stars is thus possible. In addition, as shown in Fig.~\ref{fig_empHa}, starburst regions often produce very intense nebular emission, which hinders the identification of any potential underlying stellar emission. In the case of Tol89-A, any stellar emission, if present, would be weak. It thus appears difficult to use \ha\ as an unambiguous diagnostic of VMS versus WR stars.

In view of these comparisons, we defined the following tentative method for distinguishing between star-forming regions whose UV and optical is dominated by VMS, normal WR stars, or none of them at a LMC-like metallicity:

\begin{itemize}

\item VMS: \heiiuv\ is observed in emission, is broad, and has an EW$>$3~\AA. The blue bump shows broad \heiiopt\ emission, but no or very weak \blueblue\ emission. The red bump is absent or weak, and when present, a double peak is a clear sign of VMS.

\item WR stars: \heiiuv\ is present in emission and is broad. The blue bump is present, and the blue component is clearly detected. \blueblue\ can be as strong as \heiiopt. \heiiopt\ is broad. When present, the red bump shows a broad profile.

\item no VMS and no WR stars: \heiiuv\ is weak, in absorption, or undetected, and the optical bumps are not detected.
  
\end{itemize}

The criterion for the EW of \heiiuv\ relies on the measurement for R136 and also takes results of the analysis of population synthesis models into account that we present in the next sections: the current generation of population synthesis models without VMS do not produce EW(\heiiuv)$\ga 3$ \AA.

\subsection{Empirical classification of VMS and WR-hosting spectra}
\label{s_classsources}

We then proceeded with our classification scheme to tag our sample sources. The result of our classification is given in Table~\ref{tab_classif}. In Appendix \ref{ap_class} we show the spectrum of all sources, grouped according to their classification. We are able to classify 13 sources as VMS, WR, or no VMS and no WR. In addition to R136, 2 sources clearly show evidence for VMS (J1129 and J1200). Six sources are likely dominated by WR stars, and an additional one source (J1314) is a candidate for the WR classification. Four sources do not host VMS or WR stars. For the 13 remaining sources, we are unable to provide a clear classification. In all of these cases, \heiiuv\ is detected. When the other diagnostic lines are detected but do not lead to a clear classification, the tag ``unknown'' is used. When all diagnostics including \heiiuv\ are too weak or the SNR is too low, we use the questionmark as a tag. Two out of these 13 unclear sources (NGC~3125-A1 and J1215) likely host VMS or WR stars, but the strength of \blueblue\ relative to \heiiopt\ is intermediate and does not allow us to separate VMS and WR.

Beyond R136, the two galaxies with clear evidence for VMS are J1129, previously studied by \cite{senchyna20}, who suggested the possible presence of VMS in this galaxy, and J1200, a compact star-forming galaxy from the CLASSY sample \citep{berg22}. We discuss their properties below.

We classify NGC~5253-5 as an ``unknown'' source based on the narrow width of \heiiuv. \citet{smith16} argued that this cluster hosts VMS based on the presence of \heiiuv, \ion{O}{v}~1371 and the similarity of the UV spectrum to that of R136. The optical spectrum does not show any \blueblue\ emission, which would be in line with our VMS classification scheme. However, the morphology of \heiiuv\ is rather different compared to other VMS sources, which have broader lines. The full width at half maximum of \heiiuv\ is about 600~\kms\ in NGC~5253-5, but reaches 1400~\kms\ in the average spectrum of the VMS sources presented below. Hence we refrain from assigning the VMS tag to this source. 
\citet{hc06} used template spectra of LMC WN5-6 and WC4 stars to reproduce the optical bumps observed in NGC~3125-A1. They showed that their combination of spectra was also able to produce \heiiuv. Alternatively, \citet{wofford14} argued that the intense \heiiuv\ emission can be matched if VMS are present in the cluster. In our case, we are unable to firmly distinguish between these two possibilities. The \blueblue\ component of the blue bump is present, but much weaker than the \heiiopt\ emission. The red bump is present and is broad, but relatively weak. Hence VMS, WR, or a combination of both may explain the observed spectral morphology.

\begin{sidewaystable*}[!h]
\begin{center}
\caption{Classification of the sample sources according to the presence of VMS, WR stars, or none of them. See Sect.~\ref{s_classsources} for the definition of the various classes.} \label{tab_classif}
\begin{tabular}{lcccccccc}
\hline
Source           & \heiiuv\   & \heiiuv\  &  \heiiopt \ & \blueblue  &  \civopt\ & Other &  Classification  &  Literature \\
                 & emission   & EW$>$3\AA\ &  broad     &            &           &        &                 &  \\
\hline                      
R136             &  yes       &  yes      &   yes       & weak/absent &  weak / double peak  &     & VMS             & \\
NGC~3125-A1      &  yes       &  yes      &   yes       & medium      &  weak     & \nivuva\       & VMS or WR       & WR (1), VMS (2) \\
NGC~5253-5       &  yes       &  no       &   yes       & weak/absent &  weak     & narrow \heiiuv\ & unknown         & VMS (3)  \\
Tol~89-A         &  yes       &  yes      &   yes       & medium      &  broad/strong &                & WR              & WR (4) \\
NGC~5447S        &  yes       &  no       &   ?         & weak/absent &  ?        & \heiiuv\ central abs.  & ?               &  \\
NGC~5455         &  yes       &  no       &   yes       & medium      &  broad/strong &  \nivuvb\  & WR              &  \\
NGC~5462S        &  no        &  no       &   no        & ?           &  ?        &                & none            &  \\
SIP2007-1        &  yes       &  no       &   yes       & strong      &  ?        &  \nivuva\ \&  \nivuvb & WR              &  \\
HGGK~1054        &  yes       &  no       &   yes       & strong      &  ?        &  \nivuva\ \&  \nivuvb  & WR              &  \\
HGGK~1216        &  yes       &  no       &   no        & weak/absent &  absent   &   nebular \heiiopt\ ? & unknown            &  \\
NGC~4214         &  yes       &  yes      &   yes       & absent      &  --       &   weak \heiiopt  & unknown             &  \\
NGC~4670         &  yes       &  no       &   no        & absent      &  --       &                & unknown            &  \\
Mrk~33            &  no        &  no       &   ?         & weak/absent &  weak/absent &        & none            &  \\
J0036$-$3333     &  no        &  no       &   no        & absent      &  absent   &        & none            &  \\
J0115$-$0051/SB49       &  yes       &  no       &   yes       & medium      &  broad/strong  &  \nivuva\ \&  \nivuvb & WR              & VMS? (5) \\
J0823$+$2806     &  yes       & no        &  ?          & absent      &  ?        & \heiiuv\ central abs. & unknown             &  \\
J0942$+$0928/SB80       &  yes       & no        &  yes        & medium?     &  ?        &        & ?       &  \\
J1105$+$4444     &  yes       & no        &  no         & absent      &  weak/absent &     & ?             &  \\
J1129$+$2034/SB179      &  yes       & yes       &  yes        & weak/absent &  ?        & \nivuvb & VMS             & VMS? (5) \\
J1132$+$1411/SB125     &  yes       & no        &  no         & weak/absent &  ?        & weak \heiiuv, central abs.       & none            & VMS? (5) \\
J1200$+$1343     &  yes       & yes       &  yes        & weak/absent &  ?        & \nivuva   & VMS             &  \\
J1215$+$2038/SB191      &  yes       & yes       &  yes        & medium      &  weak     & \nivuvb  & VMS or WR       & VMS? (5) \\
J1304$-$0333/SB9        &  yes       & no        &  yes        & strong      &  broad/strong  & weak \nivuvb       & WR       & VMS? (5) \\
J1314$+$3452/SB153     &  yes       & no        &  ?          & medium      &  broad    & weak \nivuva, \nivuvb  & WR?       & VMS? (5)  \\
J1428$+$1653     &  yes       & no        &  no         & absent      &  ?        &        & ?            &  \\
J1525$+$0757     &  yes       & yes       &  no         & absent      &  weak     &        & ?           &  \\
\hline
\end{tabular}
\tablefoot{Columns are source name and classification from this study and the literature. For the latter, we also give the reference: 1- \citet{hc06}; 2- \citet{wofford14}; 3- \citet{smith16}; 4-\citet{sidoli06}; 5- \citet{senchyna20}.}
\end{center}
\end{sidewaystable*}

\begin{figure}[t]
\centering
\includegraphics[width=0.49\textwidth]{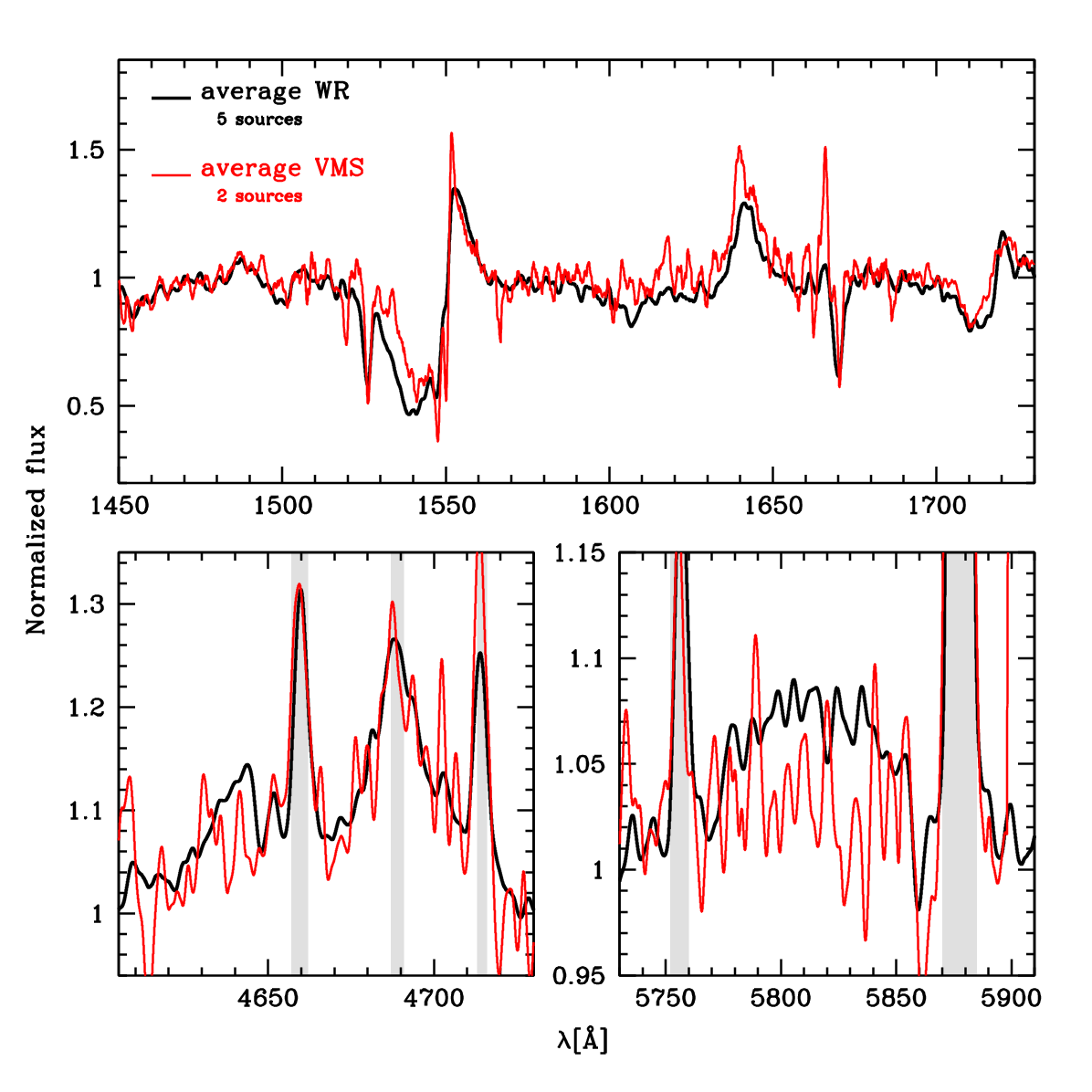}
\caption{Comparison between the average spectrum of VMS and WR sources in red and black, respectively, excluding the template sources R136 and Tol~89-A. Nebular lines are marked by gray areas.
The region above 1690~\AA\ in the red spectrum comes from J1129 alone because this region is not available for J1200.}
\label{fig_specave}
\end{figure}

Fig.~\ref{fig_specave} shows the average spectrum of the sources classified as VMS (two objects) and WR (five objects). The template R136 and Tol~89-A are not included in the average spectra. The morphology of the UV lines is rather similar, and \heiiuv\ is slightly stronger in the VMS sources.
In the optical range, the spectra are relatively noisy. WR sources show \blueblue\ emission, but this feature is somewhat weaker in VMS sources. However, the low SNR and the presence of [\ion{Fe}{iii}]~4658 nebular emission complicates the comparison. The \heiiopt\ line has a similar strength in both types of sources. The red bump has a broad emission profile in WR sources. For VMS sources, no broad emission is detected despite the low SNR. The narrow \ion{C}{iv} double peak is not detected either. Clearly, a higher SNR is required to properly investigate the morphologies of the optical bumps in VMS- and WR-dominated sources.


\subsection{UV and optical emission line strengths}
\label{s_ew}

We now examine the behavior of the main emission lines that were used for the classification in a quantitative manner. To do this,
we measured the EW of \heiiuv, the blue bump, \heiiopt,\ and the red bump using the wavelength ranges listed in Table~\ref{tab_ew1}. The EW of the \blueblue\ component can be obtained from EW(\heiiopt) and EW(blue bump). 

The EW measurements have a number of uncertainties. The main source is the low SNR in some sources. This causes difficulties in identifying the continuum, and thus in the normalization of the spectra. Noise itself introduces some bias because we work in relatively narrow wavelength ranges in which only a few data points are available. The nebular lines, especially in the blue bump, are a problem. Although they are sometimes clearly identified ([\ion{Fe}{iii}]~4658, [\ion{Ar}{iv}]~4711 see Fig.~\ref{fig_specave}) their contribution to the observed flux is difficult to constrain given the low spectral resolution of most data. This is also true of nebular \heiiopt\ when it is present. We thus estimate that our EW measurements should be taken with a $\pm$1~\AA\ uncertainty on average. 

\begin{figure*}[!ht]
\centering
\includegraphics[width=0.49\textwidth]{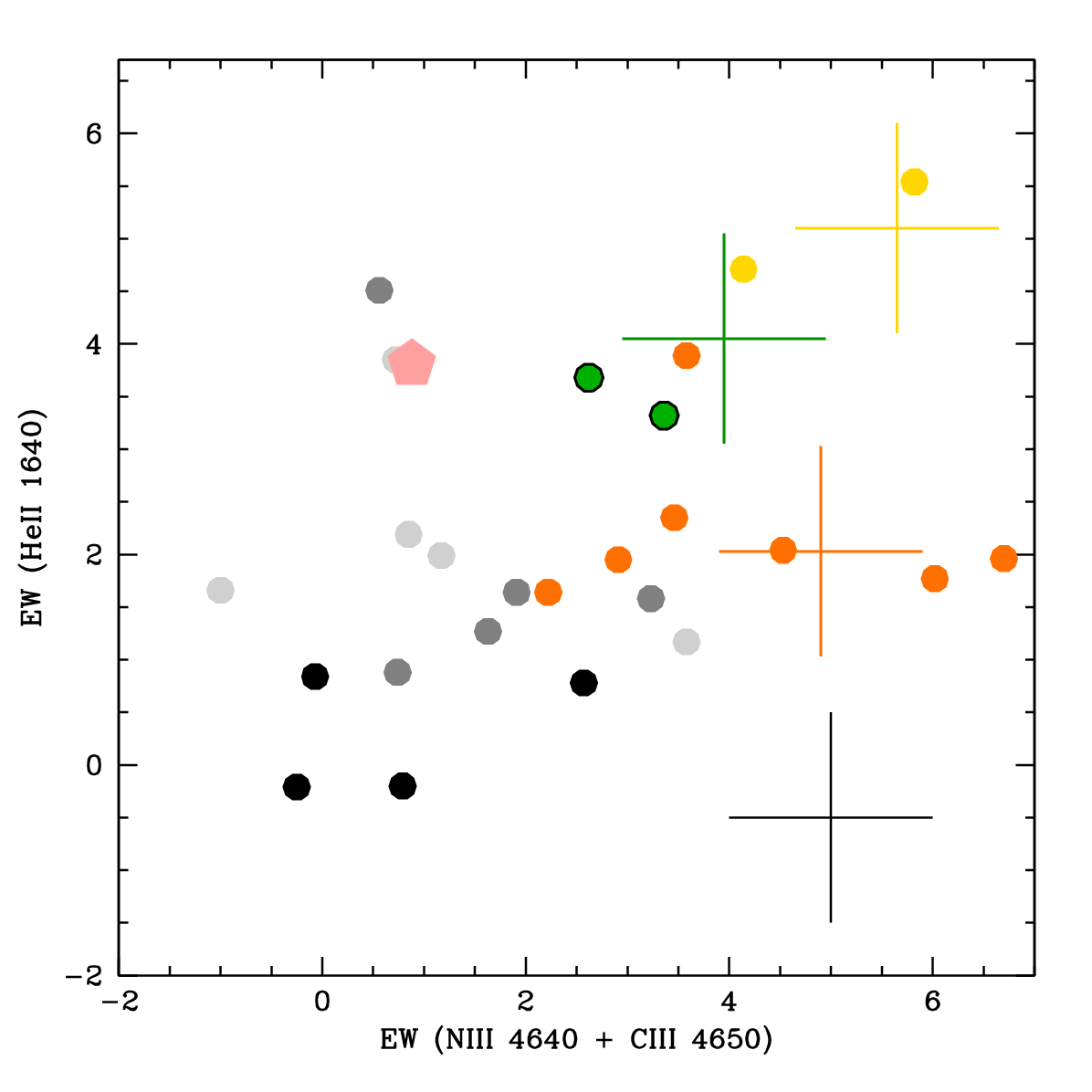}

\includegraphics[width=0.49\textwidth]{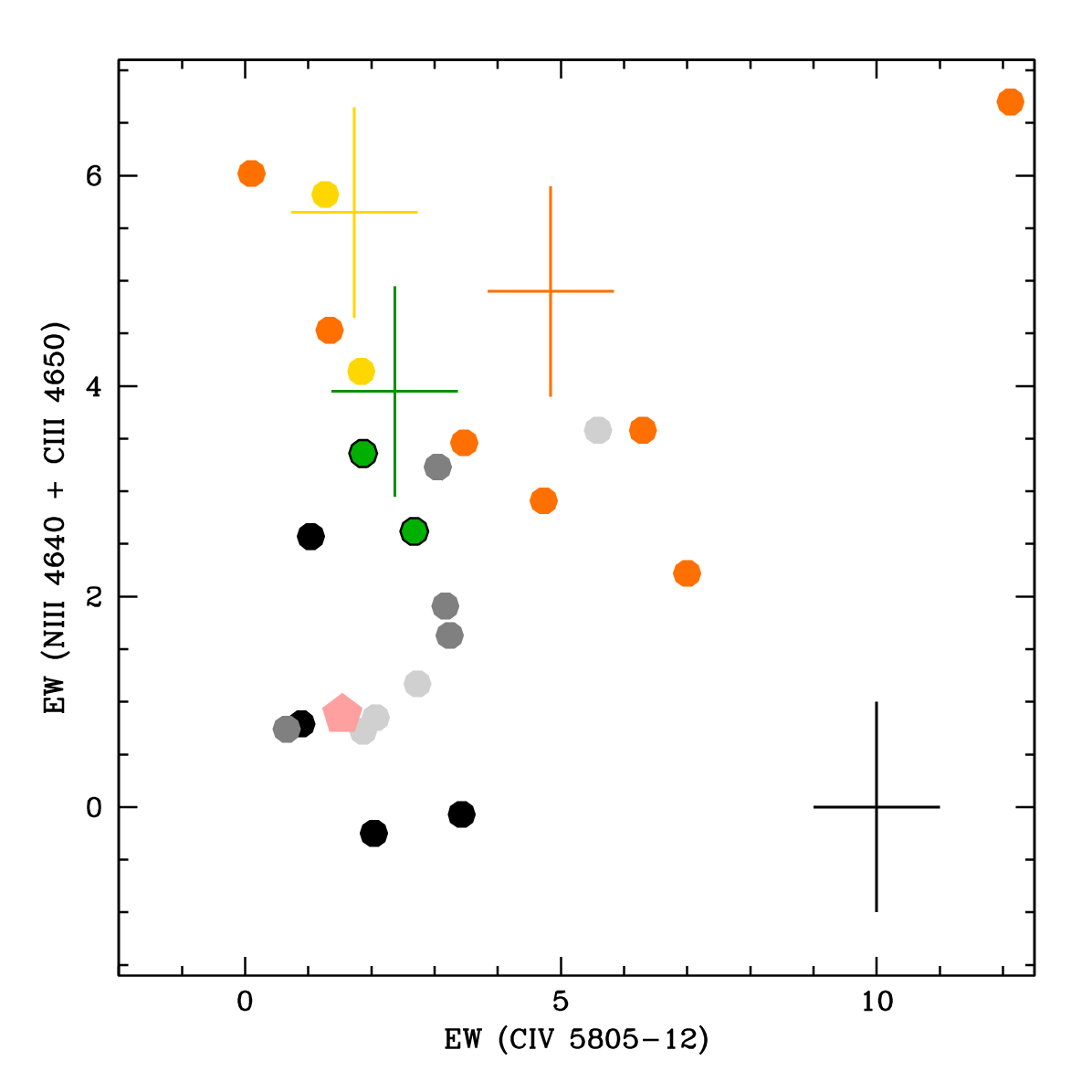}
\includegraphics[width=0.49\textwidth]{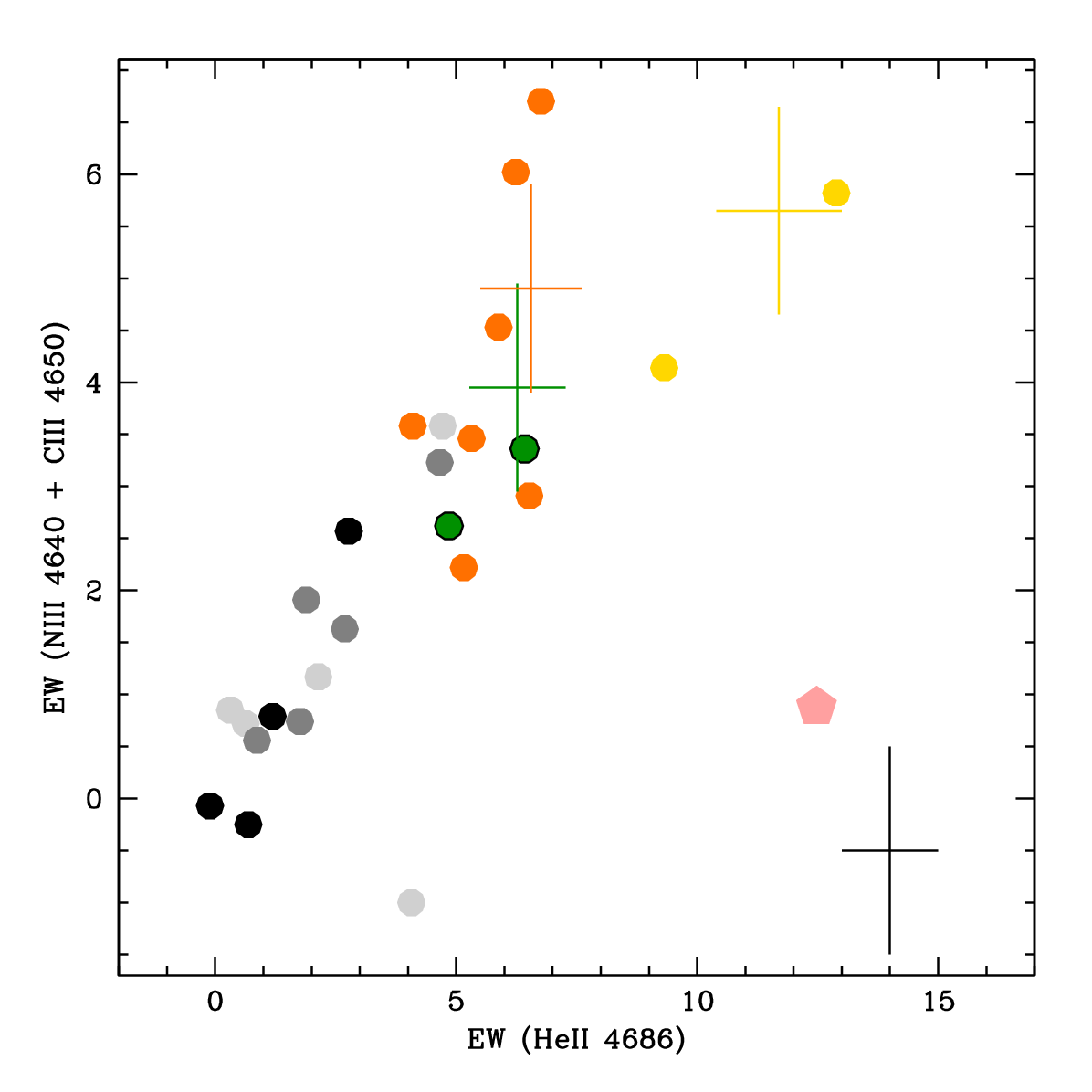}
\caption{Comparison between the EW of various lines. \textit{Top panel:} EW(\heiiuv) vs. EW(\blueblue). \textit{Bottom left panel:} EW(\blueblue) vs. EW(\civopt). \textit{Bottom right panel:} EW(\blueblue) vs. EW(\heiiopt). Circles show the sample sources. Green (with an external black circle), orange, and yellow correspond to sources hosting VMS, WR, and VMS or WR, respectively. Black symbols correspond to sources without VMS and WR stars. Dark (light) gray symbols are sources tagged as ``unknown'' (``?''). The pink pentagon is R136. The typical uncertainties on EW measurements are shown by the black cross. The green, orange, and yellow crosses correspond to the measurements on the average spectra shown in Fig.~\ref{fig_vms}, \ref{fig_wr}, and \ref{fig_vmsORwr}, respectively.}
\label{fig_ew}
\end{figure*}

With these limitations in mind, we plot EW(\heiiuv) as a function of EW(\blueblue) in the top panel of Fig.~\ref{fig_ew}. Given our classification scheme, these two features should allow us to separate the different types of sources. In practice, this is only partly true. Sources dominated by WR stars are located in the rightmost part of the diagram, 
at EW(\blueblue)$\ga~2$~\AA\ and EW(\heiiuv)$\gtrsim$~1.5~\AA. 
The VMS sources (green symbols) are located in the left part of the region covered by WR sources, with a clear overlap given the EW uncertainties. They have the same EW(\heiiuv) as R136 and a somewhat stronger \blueblue\ emission, although noise is high (see Fig.~\ref{fig_vms}).
 The sources for which a VMS or WR classification is unclear (yellow points) lie above the VMS and WR sources. The sources in which neither VMS nor WR stars are identified (black symbols) are found in the lower left corner. Finally, sources classified as unknown or tagged with a questionmark lie in between the VMS or WR sources on one side and the sources without VMS or WR on the other side. Better spectra (higher SNR and spectral resolution) would clearly help to refine the position of each source in this diagram. The EWs of the average spectra shown in Fig.~\ref{fig_vms} to \ref{fig_vmsORwr} are quite clearly separated in the top panel of Fig.~\ref{fig_ew}. In particular, the green and orange crosses, corresponding to the average VMS and WR sources, respectively, do not overlap.

In the bottom left panel of Fig.~\ref{fig_ew} we show the relation between  EW(\blueblue) and EW(\civopt). The WR sources have the largest EW(\blueblue) and show a range of EW(\civopt) that reaches the highest observed values. This is consistent with WC stars that have a strong \civopt\ emission. The WR sources with a small EW(\civopt) are likely dominated by WN stars that lack the red bump feature. The sources without an indication of WR or VMS are located in the lower left corner of the distribution of sources. The VMS sources lie in between them and the WR sources, with little to no \blueblue\ and \civopt\ emission. Here again the EWs of the average VMS and WR spectra are well separated.

The bottom right panel of Fig.~\ref{fig_ew} shows EW(\blueblue) versus EW(\heiiopt). Almost all sources follow a relation of larger EW(\blueblue) for larger EW(\heiiopt), with a sequence that extends from sources without VMS or WR to all sources classified as VMS, WR, or uncertain between VMS and WR. In this diagram, the position of the average VMS and WR sources (green and orange crosses) overlap partly. The reason is the low SNR of some of the optical spectra. For instance, J1314 shows an \blueblue\ component, but its spectrum has an absorption feature short of 4650~\AA\ (see Fig.~\ref{fig_wr}). This affects the EW measurement, which is about 3~\AA, similar to that of J1129 and J1200, the two VMS sources without a clear detection of \blueblue\ emission.
In the bottom right panel of Fig.~\ref{fig_ew}, R136 stands out as a unique object with the largest EW(\heiiopt) and almost no \blueblue\ emission. It may indeed be a peculiar object. It is also the only resolved object of our sample for which the detailed stellar content is known, however. The companion sources (see the morphologies in Sect.~\ref{ap_spatial}) could add an additional stellar continuum that would dilute the blue bump features and reduce their EWs. 
Detailed knowledge and a characterization of the stellar population at high spatial resolution is thus necessary to alleviate potential issues due to crowding.

This analysis of the EWs of several diagnostic lines we selected for our classification scheme indicates that the different types of sources can be reasonably well separated based on these features. We return to Fig.~\ref{fig_ew} in the next section when we discuss population synthesis models.

\section{Population synthesis models and comparisons with observations}
\label{s_popsyn}

We built population synthesis spectra of star-forming bursts using the same method as in \citet{mp22} (see their Appendix B). For stars with masses below 100~\msun, we used the single-star BPASS models of \citet{bpass} for a metallicity Z=0.006. To add the contribution of  VMS to these populations, we relied on the models of \citet{mp22}. The number of VMS was adopted from a Salpeter initial mass function (IMF), which we extended to 200~\msun. VMS are only present at ages younger than 2.5~Myr. For a constant star formation (CSF in the following), we simply added the contribution of individual bursts. We considered star formation up to 50~Myr. After 50~Myr, massive stars have disappeared and the UV spectra of population synthesis models are almost unchanged. In the following, we refer to our population synthesis models that include VMS as the ``VMS models''.

We compared our models to the BPASS collection\footnote{We used the models of the release 2.2, \url{https://bpass.auckland.ac.nz/}} that includes binaries because binarity is the main specificity of the BPASS project \citep{bpass}. We considered models with an upper mass limit of 300~\msun. We discuss the difference in the treatment of VMS between BPASS and our approach. The BPASS models that include binaries and stars up to 300~\msun\ were also chosen because they predict the strongest emission lines of all the BPASS collection. In addition, we considered the models of \citet{sv98}, which predict the intensity and EWs of a number of lines in young stellar populations, but not the detailed spectra. We refer to them as the SV98 models.

We compare these models with the properties of our sample sources both in terms of spectra and EWs. The latter were calculated in the same wavelength ranges for the models and observations (see Appendix \ref{ap_ew}).

\begin{figure}[t]
\centering
\includegraphics[width=0.49\textwidth]{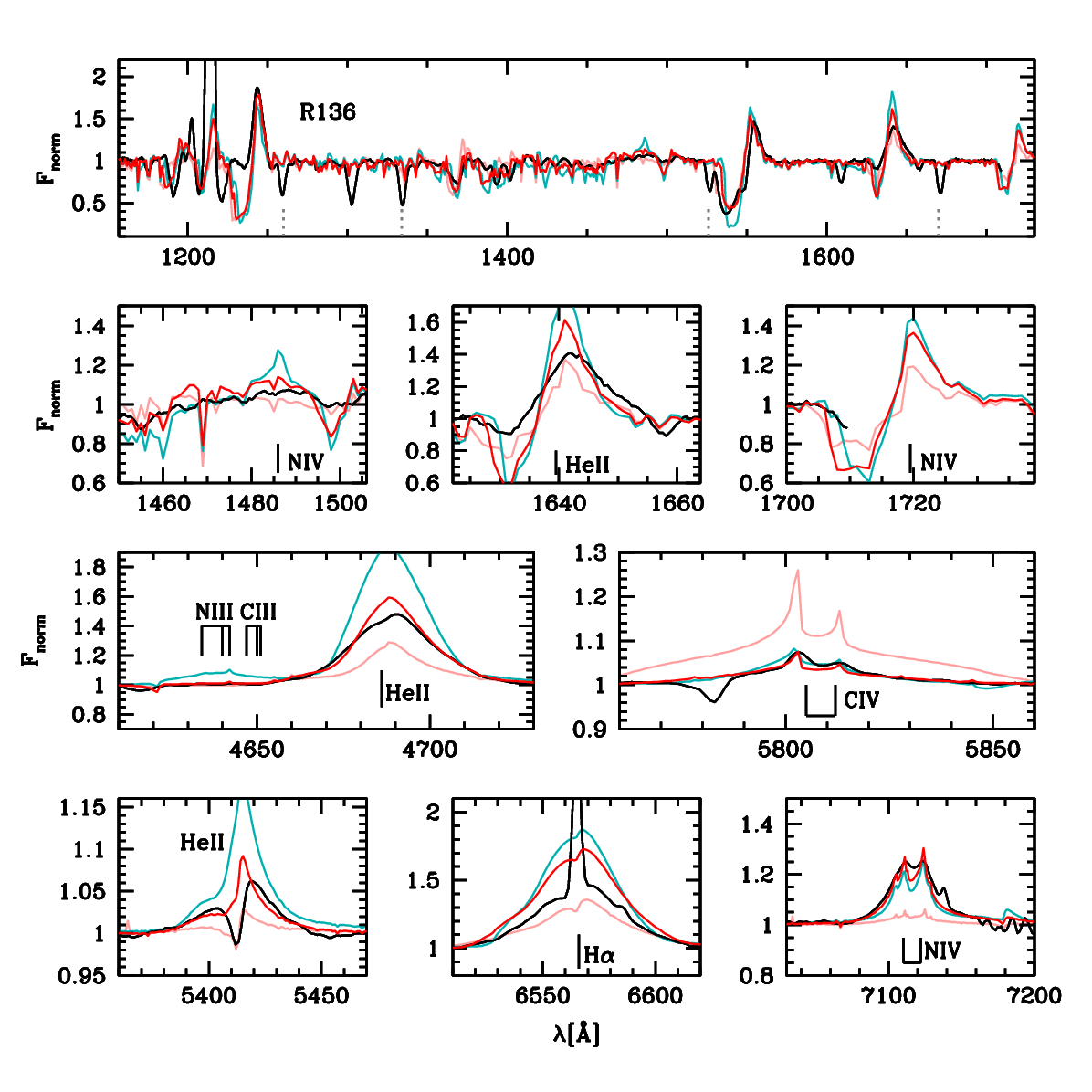}
\caption{Comparison of the normalized observed spectrum of R136 in black with population synthesis models including VMS in color. The pink, red, and cyan models correspond to an age of 1, 1.5, and 2~Myr, respectively. The main stellar lines are labeled. The gray dotted lines are interstellar features.}
\label{fig_r136}
\end{figure}

\subsection{R136 as a sanity check}
\label{s_r136}

We first investigated whether our models were able to reproduce the integrated light of R136.
The observed spectrum is shown in Fig.~\ref{fig_r136}, together with population synthesis models including VMS up to 200~\msun. We highlight the UV spectrum and several prominent optical lines that are formed in dense stellar winds. Overall, the agreement is good, especially for the 1.5~Myr model. In the UV, it reproduces the region near \nivuva\ and \heiiuv\ reasonably well. For the latter, the line morphology does not match perfectly. The blue absorption and central emission in the theoretical profile is too strong, but narrower than observed. This may be caused by an underestimate of the terminal wind velocity, as also seen in the blueward extension of the \civuv\ profile. Overall, this results in a slight underestimate of EW(\heiiuv): 2.7~\AA\ versus 3.8~\AA\ for the observed value (see also Fig.~\ref{fig_ew}). The remaining part of the UV spectrum is well accounted for, in particular, \ovuv. The blue and red bumps in the optical range are very well reproduced by the 1.5~Myr model. We note their sensitivity to age because they can become much stronger or weaker than observed in 0.5 Myr.  The blue bump is mainly caused by \heiiopt\ emission, \ion{and N}{iii}~4640 is extremely weak, if present at all for the considered ages. The \ion{He}{ii} lines at 5412~\AA\ are reproduced qualitatively, and the \ion{N}{iv} doublet near 7120~\AA\ is predicted in emission, as observed. \nvuv\ appears as a strong feature in our model, but the observed spectrum is severely affected by \lya,\ and the exact shape of the line profile is very uncertain. Our models also predict a strong stellar \ha\ emission. The 1 and 1.5~Myr models bracket the observed line profile (on top of which a narrow nebular component is seen). Interstellar lines are not included in our modeling, and the absorption part of \nvuv\ is incorrectly corrected for the broad \lya\ absorption.

We conclude that our VMS models are able to qualitatively and most of the time also quantitatively reproduce the main spectroscopic stellar features of R136 as a whole. This is also shown in Fig.~\ref{fig_ew2} with the solid red line. It represents our burst models at different ages, starting from 0~Myr near the (0,0) coordinates. Each open triangles then corresponds to an increase in age by 0.5~Myr, so that the model at 1.5~Myr is the fourth triangle of the sequence. This point lies close to the position of R136 in all figures. None of the other models shown in Fig.~\ref{fig_ew2} is able to explain the properties of R136. 

For completeness, we also compare the spectrum of Tol89-A with various models that include VMS in Fig. 7. According to these models, whether they are for a burst of star formation of CSF, the presence of VMS systematically leads to a ratio of \heiiopt\ / \blueblue\ that is never matched by any of the models that include VMS. 
This supports our classification scheme even more. Models without VMS are discussed in the next section.

\begin{figure*}[t]
\centering
\includegraphics[width=0.49\textwidth]{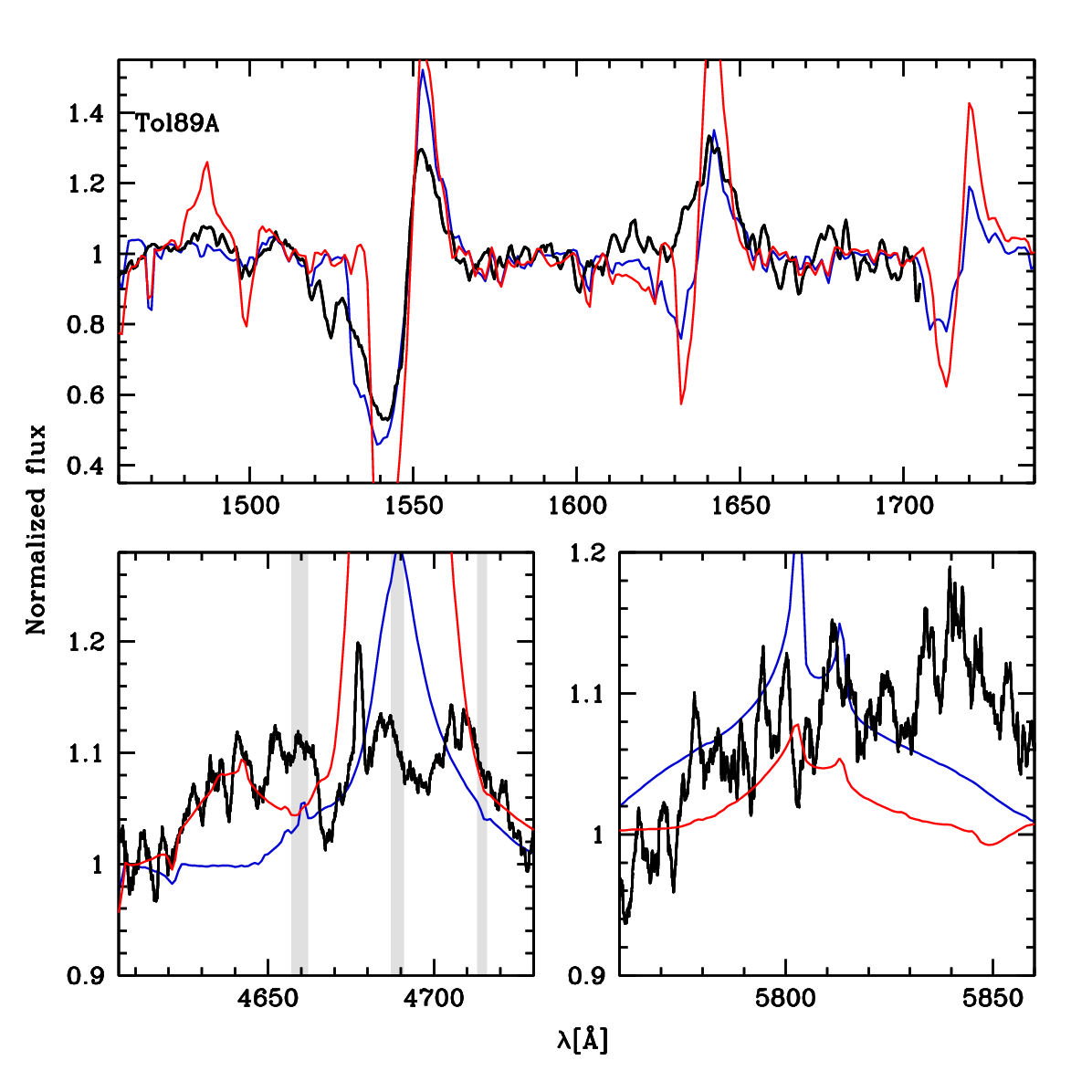}
\includegraphics[width=0.49\textwidth]{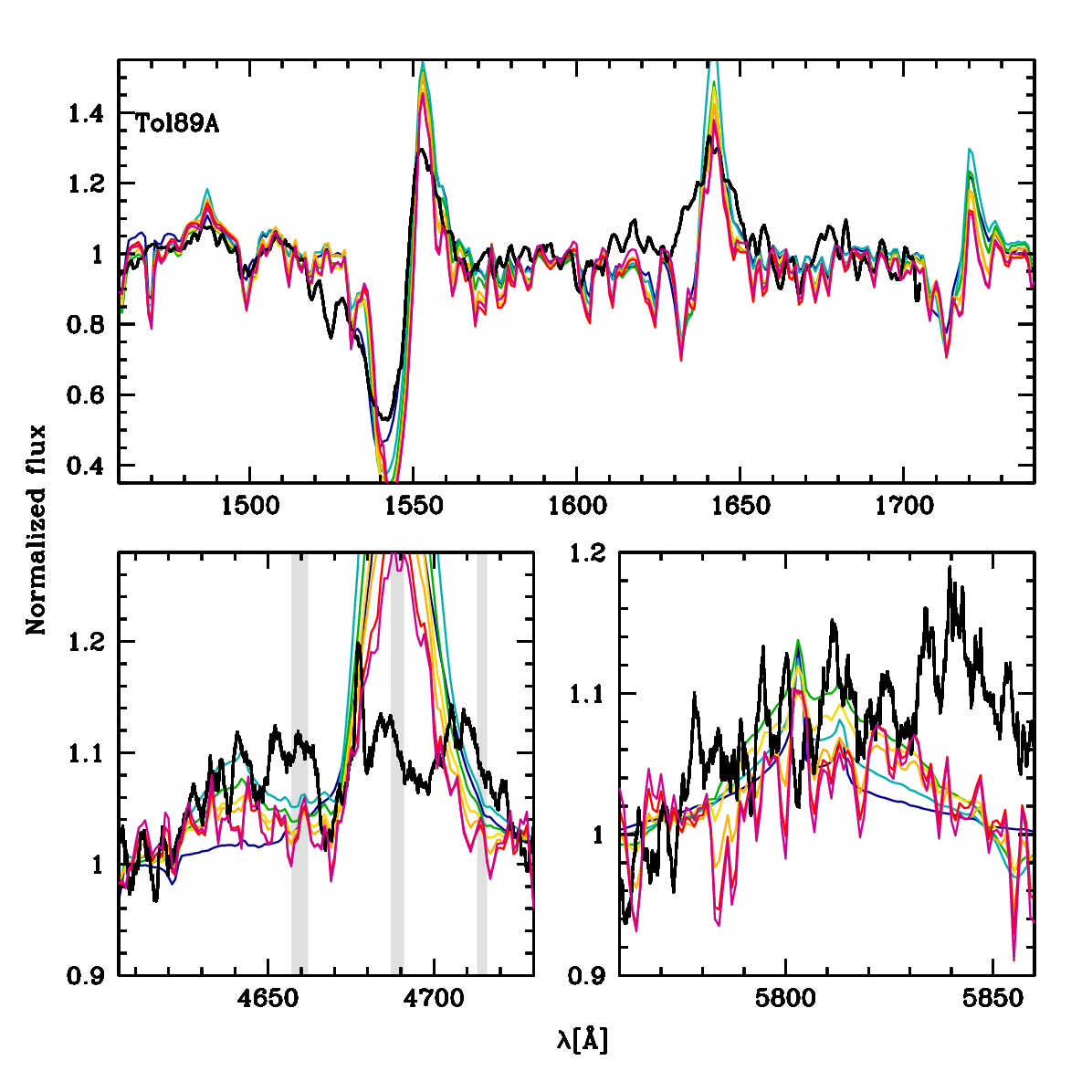}
\caption{Observed UV and optical spectra of Tol89-A in black compared to various models including VMS. \textit{Left panel:} Burst of star formation for ages 1 and 2 Myr in blue and red, respectively. \textit{Right panel:} CSF models for ages 3, 4, 5, 6, 8, and 10 Myr in blue, cyan, green, gold, orange, red, and purple.}
\label{fit_tol89}
\end{figure*}

\subsection{Successes and limitations of population synthesis}
\label{s_popshort}

\begin{figure*}[t]
\centering
\includegraphics[width=0.49\textwidth]{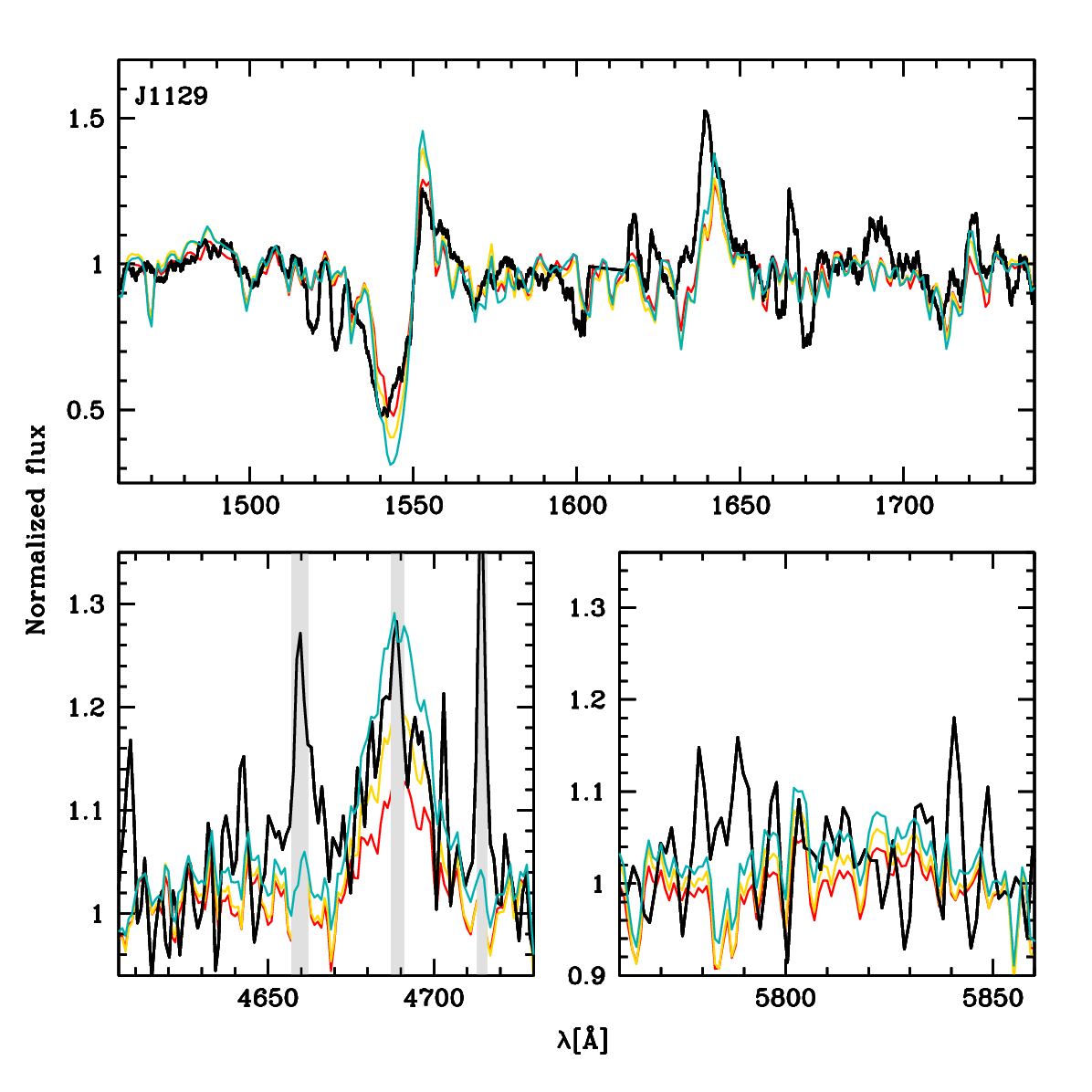}
\includegraphics[width=0.49\textwidth]{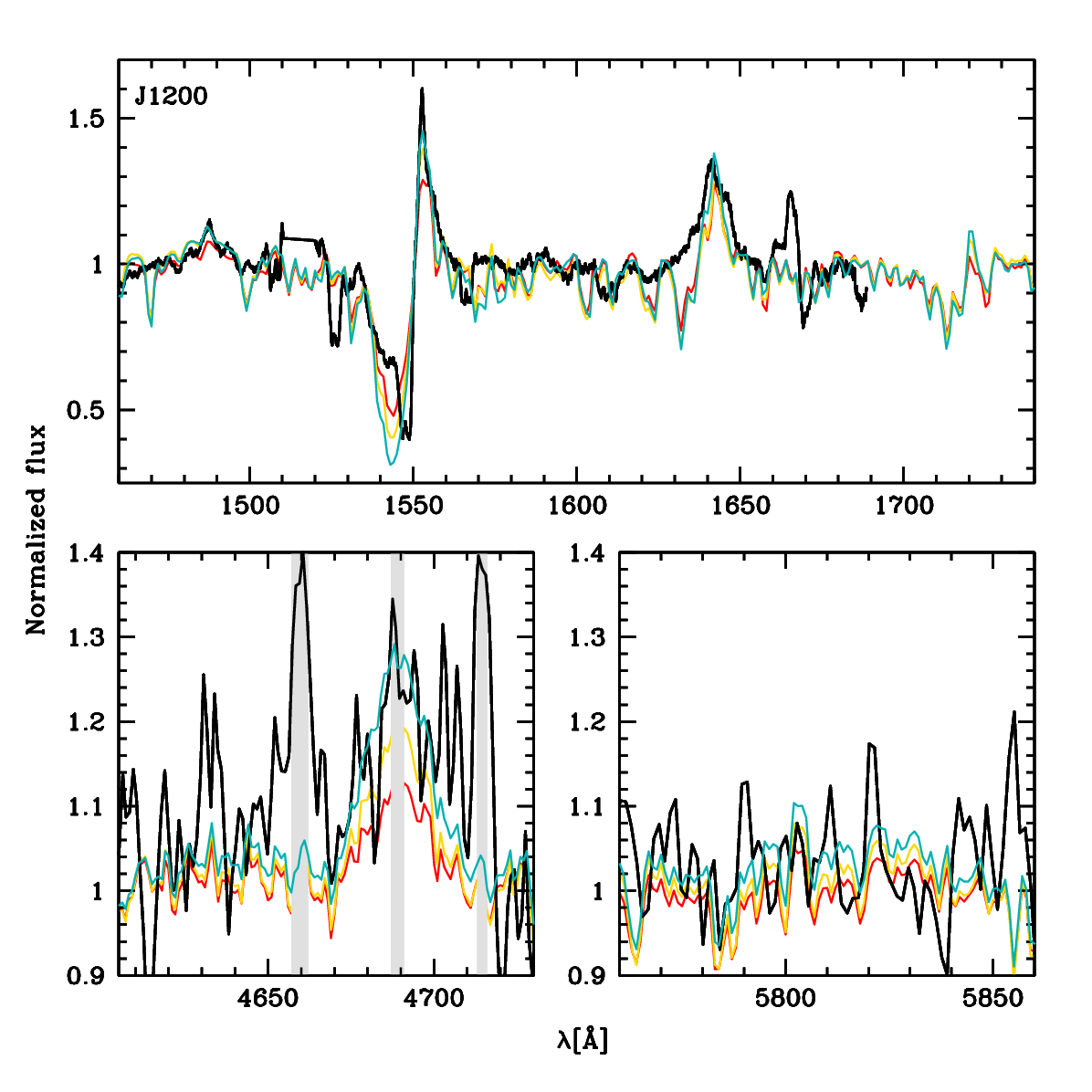}
\caption{Comparison of the normalized observed spectrum of the VMS sources in black (left: J1129; right: J1200) with CSF models that include VMS. The model shown in cyan (yellow and red) has an age of 10 (20 and 50) Myr.  The gray areas in the bottom left panel mark the position of the nebular lines of [Fe~{\sc iii}] 4658, He~{\sc ii} 4686, and [Ar~{\sc iv}] 4711.}
\label{fig_j1129vms}
\end{figure*}

We now discuss how the various models account for the properties of our sample sources in addition to R136. Fig.~\ref{fig_ew2}  compares the models to observed EWs. Focusing first on the VMS models, they cross the area in which the VMS sources are found in the EW(\heiiuv) versus EW(\blueblue) diagram, and extend to the position of the uncertain VMS or WR sources (yellow points). 
For completeness, we also considered the effect of nebular continuum emission. This additional component barely affects the UV range, but slightly dilutes the stellar spectrum in the optical range. We added nebular continuum emission following the classical approach described in \citet{osterferland}. We used their emission coefficients for hydrogen and helium, an electron temperature of 10$^4$~K, an electron density of 100~cm$^{-3}$ , and no escape of ionizing photons. A detailed description will be presented in Guibert et al. (in prep). The top panel of Fig.~\ref{fig_ew2} shows that although individual models are shifted to the left, the global shape of the line joining each series of models remains the same in the range of EWs of interest. In other words, continuum emission cannot explain sources that would have been missed by models without nebular emission.

A direct comparison between VMS models and the spectrum of the VMS source J1129 is shown in the left panel of Fig.~\ref{fig_j1129vms}. 
The models produce emission in \heiiuv\ and \heiiopt without any obvious feature at 4640~\AA\ or in the red bump, as observed. \heiiuv\ is slightly underestimated by the models, although its exact shape depends on the somewhat uncertain normalization due to the wiggles on either side of the emission line. \heiiopt\ is bracketed by the 10 and 20~Myr CSF models. \nivuvb\ is also well accounted for by these two models: this is an additional constraint compared to R136, where this feature is not covered by the observed spectrum. The right panel of Fig.~\ref{fig_j1129vms} shows the comparison of CSF models for the other VMS source, J1200. The conclusions are similar, and the  models reproduce most features relatively well. The observed spectrum is rather noisy in the blue bump region, although broad \blueblue\ emission can be excluded. 

\begin{figure*}[t]
\centering
\includegraphics[width=0.49\textwidth]{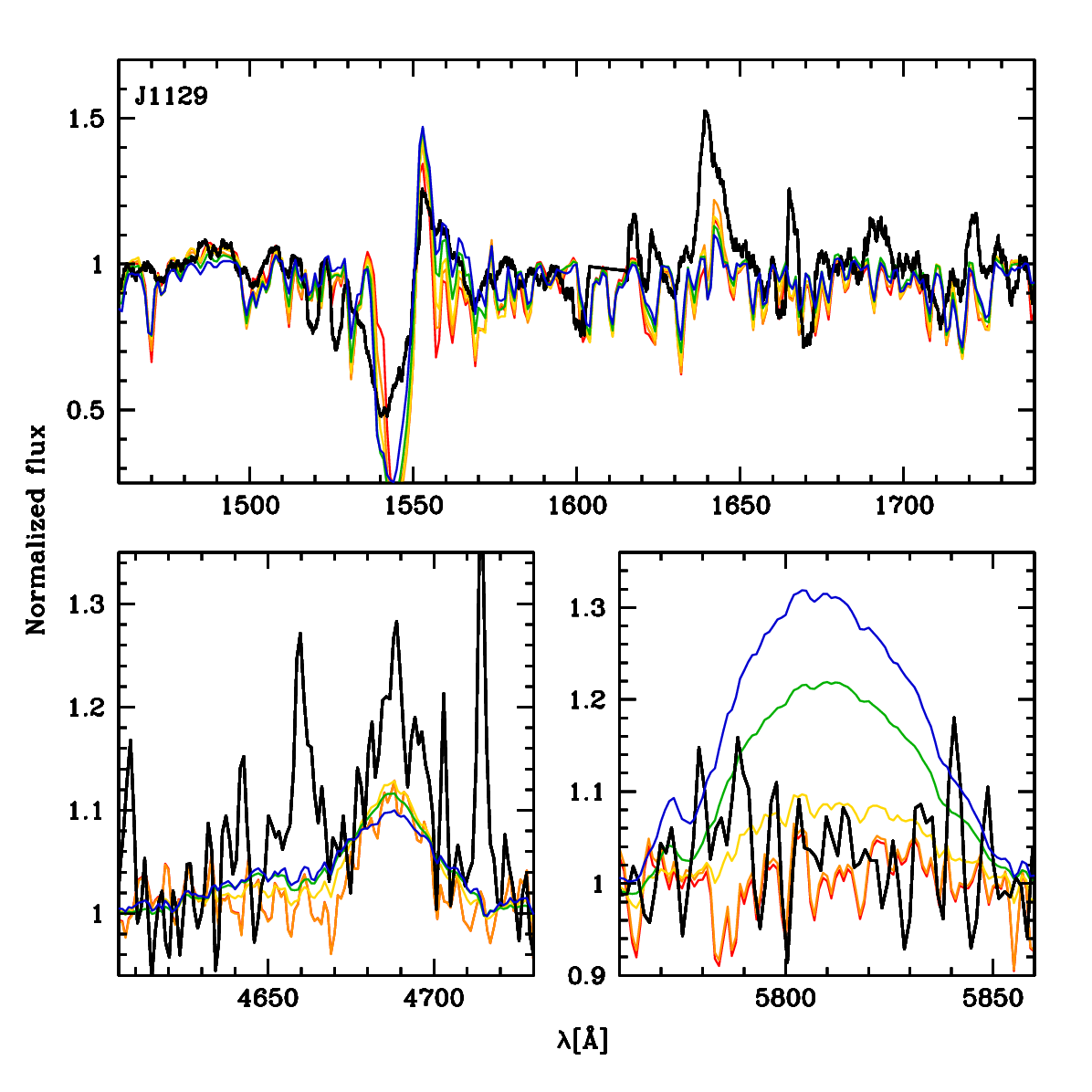}
\includegraphics[width=0.49\textwidth]{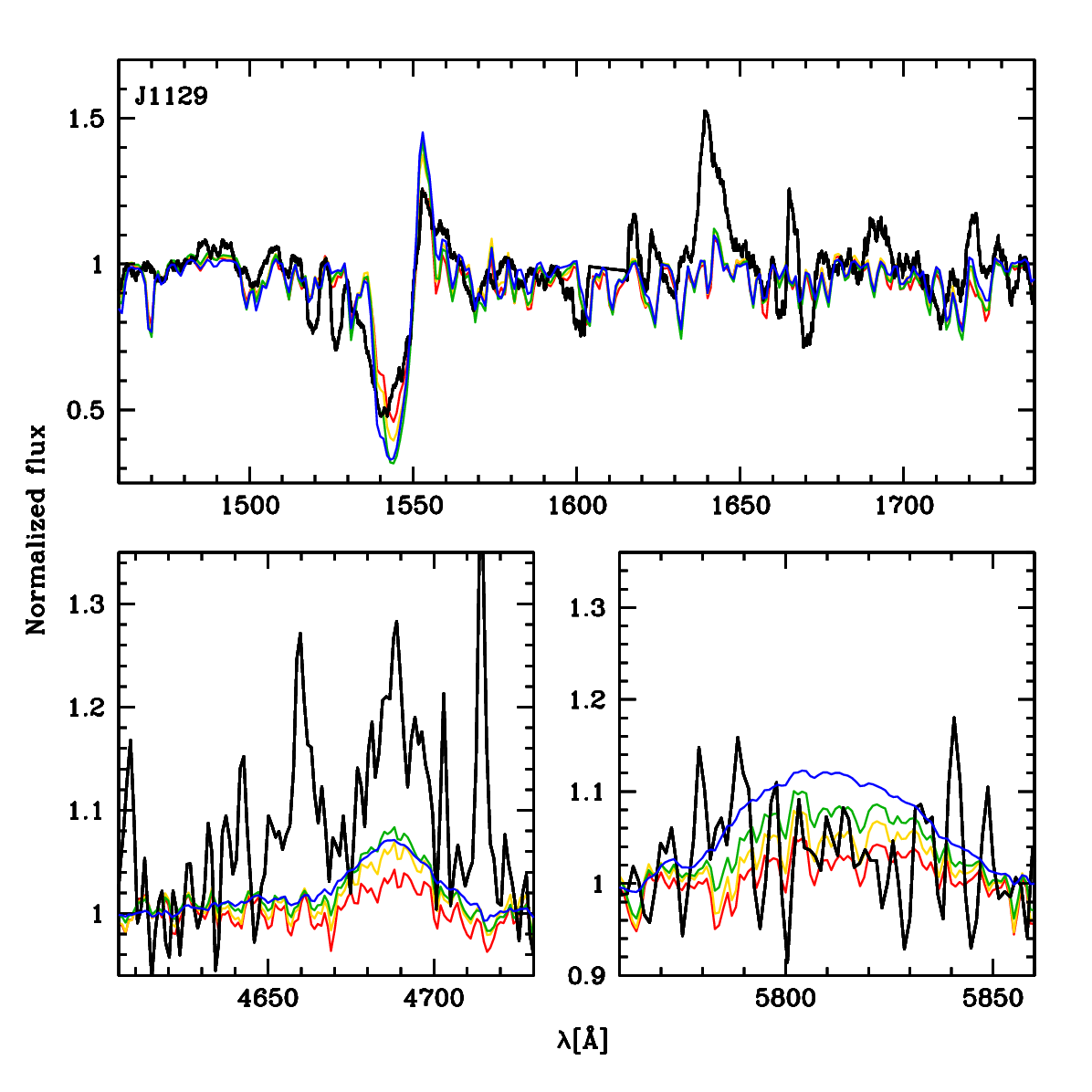}
\caption{Comparison of the normalized observed spectrum of J1129 in black with BPASS models. \textit{Left panel}: Burst models at 3, 4, 5, 6, and 8~Myr in blue, green,yellow, orange, and red, respectively. The models include binaries and an upper-mass cutoff of 300~\msun\ , as implemented in \citet{bpass}. \textit{Right panel}: CSF models at an age of 5, 10, 20, and 50~Myr in blue, green, yellow, and red, respectively.}
\label{fig_j1129bp}
\end{figure*}

The BPASS models are confined to the lower left corner of Fig.~\ref{fig_ew2}: they only account for a few sources in which we do not detect VMS or WR stars. In most BPASS models, \heiiuv\ has a negative EW, that is, it is mostly in absorption. EW(\blueblue) remains below 2~\AA. In the bottom left panel of Fig.~\ref{fig_ew2}, the BPASS models keep an EW of \heiiopt\ below the values measured in VMS or WR sources. In contrast, they reach the large EW(\civopt) seen in some WR sources. We thus conclude that the BPASS models lack some ingredients to reproduce the properties of the VMS or WR sources of our sample. This is directly visible in Fig.~\ref{fig_j1129bp}, where we compare all types of models (bursts and CSF) with the VMS source J1129. None of the BPASS models produces enough \heiiuv\ and \heiiopt\ emission. The shape of \nivuvb\ is not reproduced either. We recall that we used the BASS models with an upper mass limit of 300~\msun, that is, VMS are included at least partially (see below).

The SV98 models in the top panel of Fig.~\ref{fig_ew2} show that the burst models are almost superimposed on the VMS burst models for ages younger than 1.5~Myr. The SV98 models reach a maximum EW(\heiiuv) of about 3~\AA. Because these models did not specifically include VMS (but see Sect.~\ref{s_popdisc}), we used this value as a conservative threshold to classify sources as VMS (see Sect.~\ref{s_classif}).
The main issue regarding the SV98 models is that EW(\blueblue) remains below $\sim$1.5\AA, which is lower than that of all WR sources. Similarly to the BPASS models, the intensity of the red bump reaches the level that is observed in WR sources. This indicates a shortcoming in the modeling of the \blueblue\ feature of the blue bump.

The bottom panels of Fig.~\ref{fig_ew2} show the issue faced by population synthesis models more clearly: EW(\blueblue) is almost always underestimated by the models. The most extreme of our VMS burst models reaches the highest observed EW(\blueblue) (see Tables~\ref{tab_ew2}), but at the same time, \heiiopt\ (and \heiiuv) is much too strong. Both the BPASS and SV98 models account relatively well for the red bump, as also shown in Fig.~\ref{fig_j1129bp}, where a broad \civopt\ emission is sometimes predicted. They fail to predict strong \blueblue\ emission, however. Because \blueblue\ is produced in WN, WC, and VMS, but a strong and broad red bump is mainly caused by WC stars, this indicates a shortcoming in the treatment of WR stars, especially WN stars, in the population synthesis models.

\subsection{Potential improvements}
\label{s_popdisc}

\citet{bpass} performed quantitative comparisons between the strength of the blue and red bumps and the BPASS models for a range of metallicities and ages. Their results are shown in their Fig.~28.
They identified problems with their models when they compared their results to the sources of \citet{brinch08}: the latter have EW(blue bump)$>$8~\AA, which is not reached by any BPASS model. In addition, \citet{es09} performed direct spectroscopic comparisons between the BPASS spectra and the observed spectrum of Tol89-A. Their Fig.~7 clearly shows that the relatively strong \ion{N}{iii}~4640 emission is not produced by the BPASS models, which otherwise give a reasonable fit to all other emission lines.
The BPASS models almost all produce a net \heiiuv\ absorption (in the wavelength range we used to compute the EW), while our VMS models all have a net emission, even though stars with masses up to 300~\msun\ were included in the BPASS models. The difference is that our models include not only evolutionary tracks for VMS, but also dedicated spectra. Although VMS evolutionary tracks are considered in BPASS, the spectra used for VMS remain the same as those for normal OB and WR stars.  We attribute the lack of significant \heiiuv\ emission of the BPASS models to this limitation. By accounting for a peculiar type of evolution for very fast rotating massive stars \citep[quasi-homogeneous evolution; ][]{maeder87}, \citet{es12} were able to boost the \heiiuv\ emission of their BPASS models without VMS up to a maximum of 2.7~\AA\ at Z=0.008 (see their Fig.~2). Although significantly stronger than their standard models, this increase is insufficient to reproduce the sources with the largest EW(\heiiuv), and it remains below the threshold of 3~\AA\ we established for the VMS classification.

The SV98 models rely on empirical line luminosities of various types of WR stars, but using values preceding the recent revision of \citet{crowther23}.
For WN objects, Schaerer \& Vacca defined two classes: WNL from WN6 to WN9, and WNE from WN2 to WN4. WN5 stars were not considered. They excluded objects with low H content from the WNL subclass and excluded H-rich stars in the WNE subclass. Because VMS appear to be WN5-7h objects \citep{crowther10}, they were thus partly included in the SV98 models. This is reflected in the sanity checks that Schaerer \& Vacca performed: They compared the number of stars inferred from their line luminosities to the number of stars in 30-Dor, NGC604, and NGC~3603. They found excellent agreement. These regions all host VMS classified as WN5-7h stars. We thus conclude that the predictions of SV98 already incorporated part of the contribution of VMS, although not all of it. This explains their position in the top panel of Fig.~\ref{fig_ew2} compared to our new models that include VMS.

We showed in Sect.~\ref{s_popshort} that the strength of \blueblue\ was underpredicted by almost all population synthesis models, regardless of whether VMS were included. We describe several potential reasons for this behavior. First, our VMS models are computed for a given chemical composition (1/2.5 \zsun; \citealt{mp22}). VMS quickly reach the CNO equilibrium, so that the nitrogen content at the stellar surface reflects the initial carbon content. For higher initial C/H, the \ion{N}{iii}~4640 emission would be stronger. Second, the models of \citet{mp22} did not cover the entire evolution of VMS. The latest phases are missing. Although very short (a few 10$^5$~yr), these phases correspond to states in which the stars are more chemically processed, with barely any hydrogen left. The spectroscopic appearance should thus be similar to that of normal WR stars that show strong \blueblue\ emission. The impact of these short (but bright) phases on population synthesis models remains to be studied.
Additionally, the BPASS models consider WR spectra only when the surface hydrogen mass fraction of the stellar models is lower than 0.4. The analysis of LMC WN stars by \citet{hainich14} indicated that in about one-third of them (30 out of 107),  X$_{H}$ is equal to or higher than 0.4. These objects are also the most luminous (see Fig.~7 of Hainich et al.). BPASS might therefore underestimate the contribution of these H-rich WN stars to the optical bumps, and to the \ion{N}{iii}~4640 emission in particular. The PYPOPSTAR models developed by \citet{hrpypopstar} have the same limitation: According to \citet{molla09}, WR spectra are used only for stars with a surface hydrogen mass fraction smaller than 0.3. This leaves out a number of true WR objects in the LMC. We state again that \citet{es09} reported an example for which the BPASS models missed the emission in \blueblue, likely because of the treatment of WR stars. 
Finally, \citet{hc06} reproduced the optical spectrum of NGC3125-A1 using empirical spectra of normal WR stars. Their approach did not rely on population synthesis, but on coadding individual spectra of WN5-6 and WC4 stars until a match to the observed profiles of the blue and red bumps was found. They succeeded in reproducing the blue bump. Hence, the treatment of normal WR stars in population synthesis models appears to be key to solving the problem of underpredicted \blueblue\ emission.

\section{Discussion}
\label{s_disc}

\begin{figure}[t]
\centering
\includegraphics[width=0.49\textwidth]{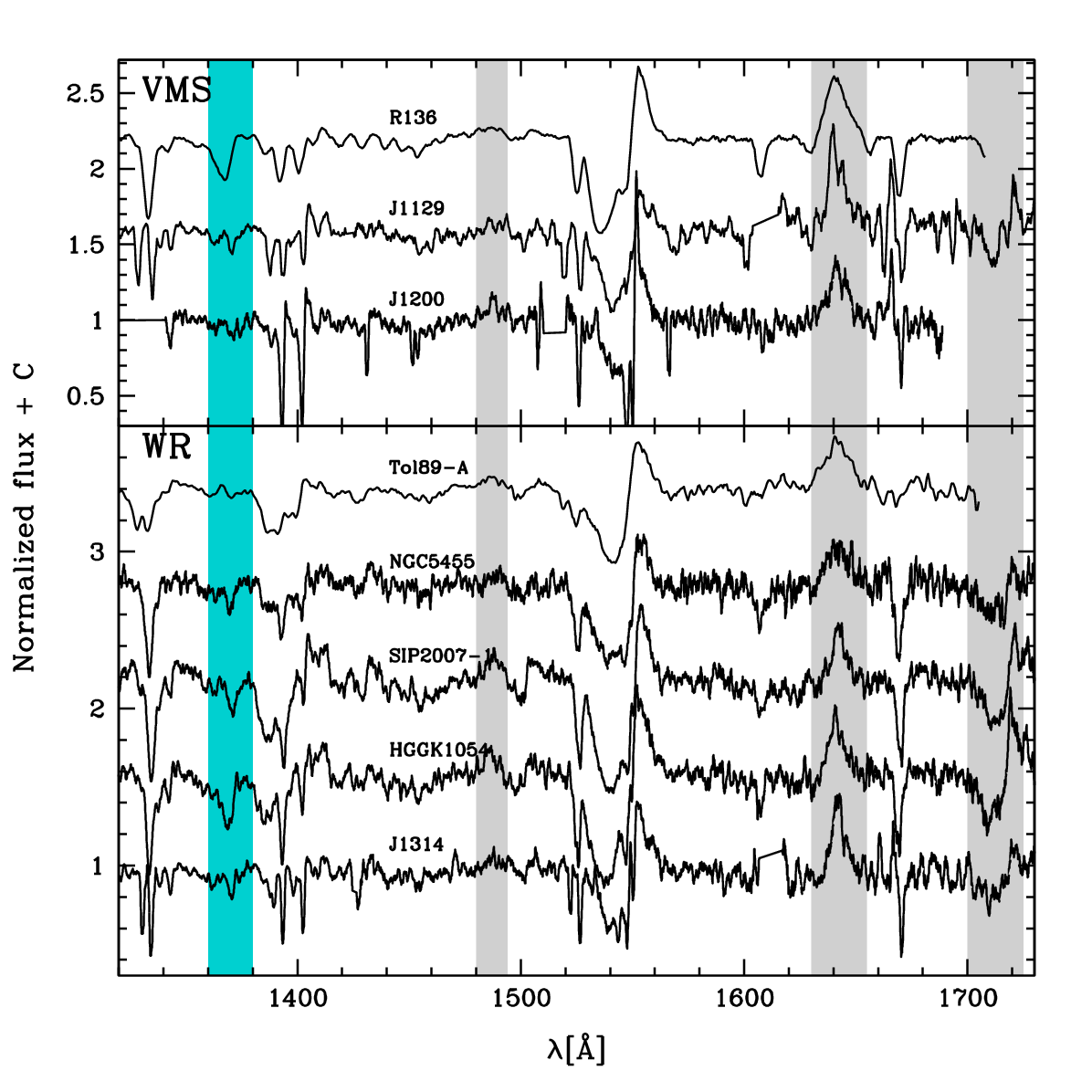}
\caption{UV spectrum of galaxies and/or regions hosting VMS (top) and WR (bottom) stars. The gray areas show the position of \nivuva\ and \heiiuv. The cyan area shows the \ovuv\ feature.}
\label{fig_ov}
\end{figure}

\subsection{On \ovuv\ as a VMS indicator}

In the first part of this paper, we proposed an empirical scheme to infer VMS and WR stars in young star-forming regions. \citet{wofford14,wofford23} and \citet{smith16} used \ovuv\ as a complementary diagnostic for VMS in NGC3125-A1 and NGC5253-5, respectively. This feature requires high ionization in stellar atmospheres and thus high effective temperatures. The earliest O stars display strong \ovuv\ profiles, with a P-Cygni shape in the giant and supergiant luminosity class \citep{wnp85}. In addition, oxygen must be present in sufficient number for \ovuv\ to develop. Because oxygen is depleted in the phases that precede helium burning, its surface abundance decreases from the main sequence to the WR phases. \ovuv\ is thus present in the earliest phases of stellar evolution of the most massive stars. VMS fit this category.

Given the required conditions to form \ovuv,\ some WR stars can also produce it. \citet{niedroch94} compiled low-resolution \textit{\textup{International Ultraviolet Explorer}} spectra of WN and WC stars and showed that \ovuv\ is present in WC5-9 objects (see their Figs.~4 and 11d). Some WN stars also display it, although the feature is less common in that category because the oxygen abundance is lower. 

In Fig.~\ref{fig_ov} we show the UV spectrum of the sources we classified as VMS hosts (top) and WR hosts (bottom). 
\ovuv\ is clearly identified in the integrated spectrum of R136, which confirms that regions with VMS show this feature. 
\cite{crowther16} showed that \ovuv\ is present both in VMS and non-VMS stars in R136.
\ovuv\ is also present in the other two VMS sources, J1129 and J1200, but the intensity is weaker. When we inspect the morphology of the WR-dominated sources, \ovuv\ is detected in most sources. SIP2007-1 and HGGK~1054 show strong \ion{O}{v} lines. At the same time, we recall that the morophology of their blue bump is close to that of Tol89-A, indicating the presence of WR stars. NGC~5455 has both \ovuv\ and a broad \civopt\ emission (see Fig.~\ref{fig_wr}), which is consistent with WR stars. Based on these empirical comparisons, we thus refrain from including \ovuv\ in our classification scheme because like other UV lines, it may be present in both VMS- and WR-dominated sources.

\subsection{Comparison with other studies of VMS populations}

\citet{senchyna20} determined whether population synthesis models can consistently reproduce the stellar and nebular features of star-forming regions. They relied on the Charlot \& Bruzual models \citep{plat19} that incorporate VMS up to 350~\msun\ based on the evolutionary tracks of \citet{chen15}. They did not use dedicated VMS synthetic spectra, but relied on the spectral libraries of WR stars of \citet{sander12} and \citet{todt15}. They ran photoionization models with spectral energy distributions (SEDs) from their population models as input and investigated whether the predicted nebular lines reproduced the observed features. By doing this and focusing on nebular lines alone, they found that the models leading to the best agreement for nebular lines produced stellar UV lines of the underlying population that were too weak. Adding stellar UV lines to the global fit (in addition to nebular lines) improved the fit of stellar UV lines as expected, but at the cost of increasing the metallicities. At the same time, the ages were on average 7~Myr younger than when nebular lines alone were considered. When stellar lines were included in the study, the WR bumps in the optical spectrum of the underlying population were too strong (their Fig 7). Senchyna et al. concluded that the discrepancy can be resolved by considering an excess of VMS produced by stellar mergers at young ages. These stars could boost stellar wind lines and relax the need for a higher metallicity. However, this hypothesis was not tested. 

The source for which the stellar lines in the UV and optical show the largest discrepancy in the sample of \citet{senchyna20} is SB179, J1129 in our case. Their model underestimates \heiiuv\ and at the same time overestimates the WR bumps. Interestingly, J1129 is one of the two sources that we classify as VMS. We showed in Fig.~\ref{fig_j1129vms} that a CSF model at  about 20~Myr leads to a good fit of both the main UV lines and the optical WR bumps. 
Another source with large discrepancies in the sample of Senchyna et al. is SB49, J0115 for us. While Senchyna et al. were able to reproduce the UV lines, the blue bump was severely overpredicted and the red bump was simultaneously too weak. We classify J0115 as a WR source, and we argue that taking  not only the UV but also optical lines into account is equally important to assess the nature of the populations in star-forming regions. As discussed in Sects.~\ref{s_popshort} and \ref{s_popdisc}, an improvement in the inclusion of WR stars in population synthesis models will help us to better characterize the problematic sources in the sample of Senchyna et al..

At present, only a few studies have claimed that VMS are present in young stellar populations. \citet{wofford14,wofford23} and \citet{smith16} argued that the strong \heiiuv\ emission together with \ovuv\ was a clear signature of VMS in NGC~3125-A1 and NGC~5253-5, respectively. In our study, we were not able to confirm that VMS dominate the UV spectra of these two star-forming regions. 
\citet{leitherer18} reported intense \heiiuv\ emission of cluster SSC-N in II~Zw~40: its EW$=7.1$~\AA\ is similar to NGC~3125-A1 (using the measurement from their Fig.~10). 
The authors discussed the lack of VMS models in population synthesis to explain that strong \heiiuv\ is not reproduced by Starburst99 spectra. \citet{mestric23} reported strong UV lines in the young cluster Sunburst at a redshift $z=2.37$. Using an independent age estimate from \citet{chisholm19} (2.9-3.6 Myr), they favored VMS as the likely sources of these intense UV features. In our study, we were able to firmly classify only  
two sources (besides R136) as VMS hosts based on the spectroscopic similarity to R136. Together with the other investigations listed above, this indicates that VMS are difficult to identify.
On the other hand, our approach is conservative, and the presence of VMS in more of the sources studied here is not excluded. For example, ten sources have unknown or uncertain classification, half of them because the quality of the spectra is insufficient (see Table \ref{tab_classif}).

Some recent work also excluded the presence of VMS.
\citet{mayya23} studied the ionizing properties of 32 \ion{H}{ii} regions of the ring of the Cartwheel galaxy. They reported that Charlot \& Bruzual population synthesis models \citep{plat19} in the WR phase can explain nebular line intensities, especially \heiiopt. However, they did not detect the blue bump emission in any of the sources, in spite of the expected presence of WR stars. They explained this behavior by the relatively low metallicity of the ring region: their estimate is 12+log(O/H)=8.19$\pm$0.15. They also investigated whether VMS might be hidden because of the low metallicity, but concluded that they would have produced sufficiently large EW(\heiiopt) to be detected if they were present. They thus concluded that VMS were absent from their sample of \ion{H}{ii} regions.

\subsection{Properties of VMS-hosting regions and prospects for VMS searches at other metallicities}

 Only very few objects clearly show VMS. It is therefore difficult to determine the conditions that are required to form these stars. 
The three objects that we firmly classify as VMS hosts, however, share several properties that might be indicative. 
All objects are young (as indicated, e.g., by their high \hb\ equivalent widths; see Table~\ref{tab_ew2}), show a high specific star formation rate (sSFR $\sim (10-300)$ Gyr$^{-1}$),
and are compact, at least in the UV \citep[see][]{cignoni15,senchyna17,berg22}. 
Furthermore, J1200 was selected by CLASSY as an object with a high electron density ($n_e > 400$ cm$^{-3}$), which is very rare at low redshifts
\citep{berg22}. This indicates that intense star formation occurs over short timescales and in fairly compact and/or dense regions in these VMS-hosting objects. These conditions could indeed favor the formation of relatively massive star clusters and VMS within them \citep[e.g.,][]{Krause2020}.
On the other hand, we do not know if all objects with similar relatively extreme conditions host VMS, that is,\ whether these conditions would be sufficient to predict the formation of VMS. Future work, both observationally and from simulations, is clearly needed to address these questions.

Our study and most of those recalled above focused on sources with metallicities close to the metallictiy of the LMC. As explained in Sect.~\ref{s_2}, this is driven by the availability of a unique template: R136. The nature of the individual sources and the integrated light of that cluster are both known, and its VMS content has been studied in detail \citep{crowther10,crowther16,besten20}. There is no reason to think that VMS exist only at LMC metallicity, however. The brightest members of NGC~3603 in the Milky Way easily enter this category, with initial masses of about 150~\msun\ \citep{crowther10}. In the Arches cluster, the current most massive stars have masses just above 100~\msun\ \citep{arches}. The age of the cluster, 2.5~Myr, likely implies that more massive stars have already disappeared, however. Conversely, VMS probably form at lower metallicity as well. 

If VMS and WR stars are difficult to separate in the UV, it is mainly because they show the same strong lines that form in their winds. These winds are radiation driven, and the associated mass-loss rates weaken at lower metallicity \citep{krticka17,bjorklund21,sander20a}. The predictions of \citet{sander20b} for classical WR stars indicate that the onset of VMS winds caused by the proximity to the Eddington limit is pushed at a higher Eddington factor for lower-Z stars (see their Fig~11). If this predictions remains valid for H-rich WR stars such as VMS, this would imply that the strong emission lines that are seen in VMS at LMC metallicity would appear only in the latest phases of low-Z VMS evolution. VMS would thus be more difficult to identify in the integrated light of young starbursts because their strong lines would be visible for a shorter time of their entire evolution. 

Unfortunately, the prospect of identifying individual VMS in low-metallicity galaxies of the Local Group is dim. In the SMC, the most massive stars reach no more than about 60~\msun\ \citep{castro18b,ramachandran19,mp21}. The current star formation rate of the SMC is about one order of magnitude lower than that of the LMC, and it is comparable to or higher than that of other Local Group galaxies \citep[see Fig.~11 of][]{crowther19}. If VMS form preferentially in regions with a high star formation rate (which is not established), then Local Group galaxies are not the best candidates to host them. Advances in the understanding of VMS at any metallicity will thus likely rely on studies of unresolved star-forming regions. It is therefore important to properly incorporate VMS in population synthesis models, and to solve the shortcomings of the integration of normal WR stars in these models. These improvements will be the topics of subsequent publications.

\section{Conclusion}
\label{s_conc}

We have studied the presence of very massive stars ($M > 100$ \msun) in local star-forming regions. We selected starbursts with metallicities close to that of the LMC (12+log(O/H)=8.3) to compare them with templates and populations synthesis models available at that metallicity. Our sample is composed of 27 star clusters and star-forming regions with HST UV and optical spectroscopy. 
 
To firmly establish the presence of VMS from integrated spectra of star-forming regions, we have shown that VMS and normal WR star must be clearly distinguished.
This is not feasible with UV data alone, unless independent constraints on the age of the populations are available. VMS are present during the first 2-3~Myr of a starburst, while WR stars appear after 3~Myr and are seen at ages up to 5-6~Myr. In both cases, broad and strong \heiiuv\ emission is observed, sometimes accompanied by \nivuva\ and \nivuvb\ emission and P-Cygni lines. The optical WR bumps are well suited to separate VMS-dominated from WR-dominated sources. In the former, the blue bump is almost exclusively formed by a broad \heiiopt\ emission, while in the latter, the \blueblue\ component is present, sometimes as strongly as \heiiopt. The red bump can be present in some WR-dominated sources and is broad. In VMS-dominated sources, it is either absent or present in the form of a narrow double peak. These spectroscopic characteristics were mainly drawn from the comparison of the integrated spectrum of R136 (in which only VMS are present) and Tol89-A (in which the presence of WN and WC stars was inferred).

Based on these empirical spectroscopic criteria, we classified our sources into  VMS-dominated (two objects), WR-dominated (five objects and one candidate), or hosting no VMS and no WR (four objects). For two additional objects, VMS or WR are present, but we cannot favor one type of source over the other with the current data. For the remainder of the sample, the data do not lead to a clear classification, partly because the quality (SNR) of the spectra is insufficient. By construction, our approach provides a conservative classification of sources hosting VMS, and establishing the presence of VMS from integrated spectra of star-forming regions or galaxies is not trivial because their lifetimes are short and there are fewer of them than in the total stellar population.

We measured the EW of \heiiuv\ and of the components of the blue and red bumps for all sources. We showed that sources can be relatively well separated into diagrams when these measurements are combined (e.g. EW(\heiiuv) versus EW(\blueblue)). 

We compared the predictions of populations synthesis models with the measured EW of the sample sources, with special emphasis on the BPASS models that include binaries and VMS up to 300~\msun. These models are unable to produce significant \heiiuv\ emission (i.e., EW(\heiiuv) remains below 1~\AA), in contrast to the VMS- or WR-dominated sources, for which EW(\heiiuv) can reach almost 6~\AA. We attribute this failure to the use of synthetic spectra, which are not appropriate for VMS. Incorporating the stellar models of \citet{mp22}, we showed that the observed EW(\heiiuv) are well reproduced. This stresses the need to use dedicated evolutionary models, atmosphere models and synthetic spectra for VMS.

In the course of comparing predicted versus observed EWs, we found that no model (BPASS, SV98, and our own models) is able to correctly reproduce the intensity of the \blueblue\ emission. At the same time, the red bump is well accounted for. This highlights shortcomings in the inclusion of WN stars in current population synthesis models.

Further optical spectroscopy is required to investigate the presence of VMS in several of our sources. In particular, high SNR data are necessary to better quantify the difference in the morphology of the blue bump in VMS- and WR-dominated sources.
In spite of these limitations, our finding of two clear examples of VMS-dominated sources (J1129 and J1200) and earlier studies revealing the possible presence of VMS in several other \hii\ regions or galaxies \citep{wofford14,smith16,senchyna17,senchyna20,mestric23} showed that VMS are not an exotic phenomenon and that these stars are likely present in a fraction of star-forming regions.
The conditions under which they appear have to be studied with larger samples. A wider range of metallicities must clearly be explored, with potential impact for the study of the first starbursts at very high redshift. This requires the development of dedicated VMS models and will be the topic of a subsequent publication.

\section*{Acknowledgments}

We thank an anonymous referee for critical reviews of the manuscripts. We thank Linda Smith, Fabrizio Sidoli, Paul Crowther and Cesar Esteban for sharing data on NGC5253-5, Tol-89, R136 and \hii\ regions in M101 respectively. We thank the ESO support, especially John Pritchard, for help with the ESO pipelines. This research is partly based on observations made with the NASA/ESA Hubble Space Telescope obtained from the Space Telescope Science Institute, which is operated by the Association of Universities for Research in Astronomy, Inc., under NASA contract NAS 5–26555. The present study is partly based on data obtained from the ESO Science Archive Facility.

\bibliographystyle{aa}
\bibliography{vms_local}


\begin{appendix}
\label{ap_tab}

\section{Spatial morphology of the sources}
\label{ap_spatial}

\begin{figure}[t]
\centering
\includegraphics[width=0.49\textwidth]{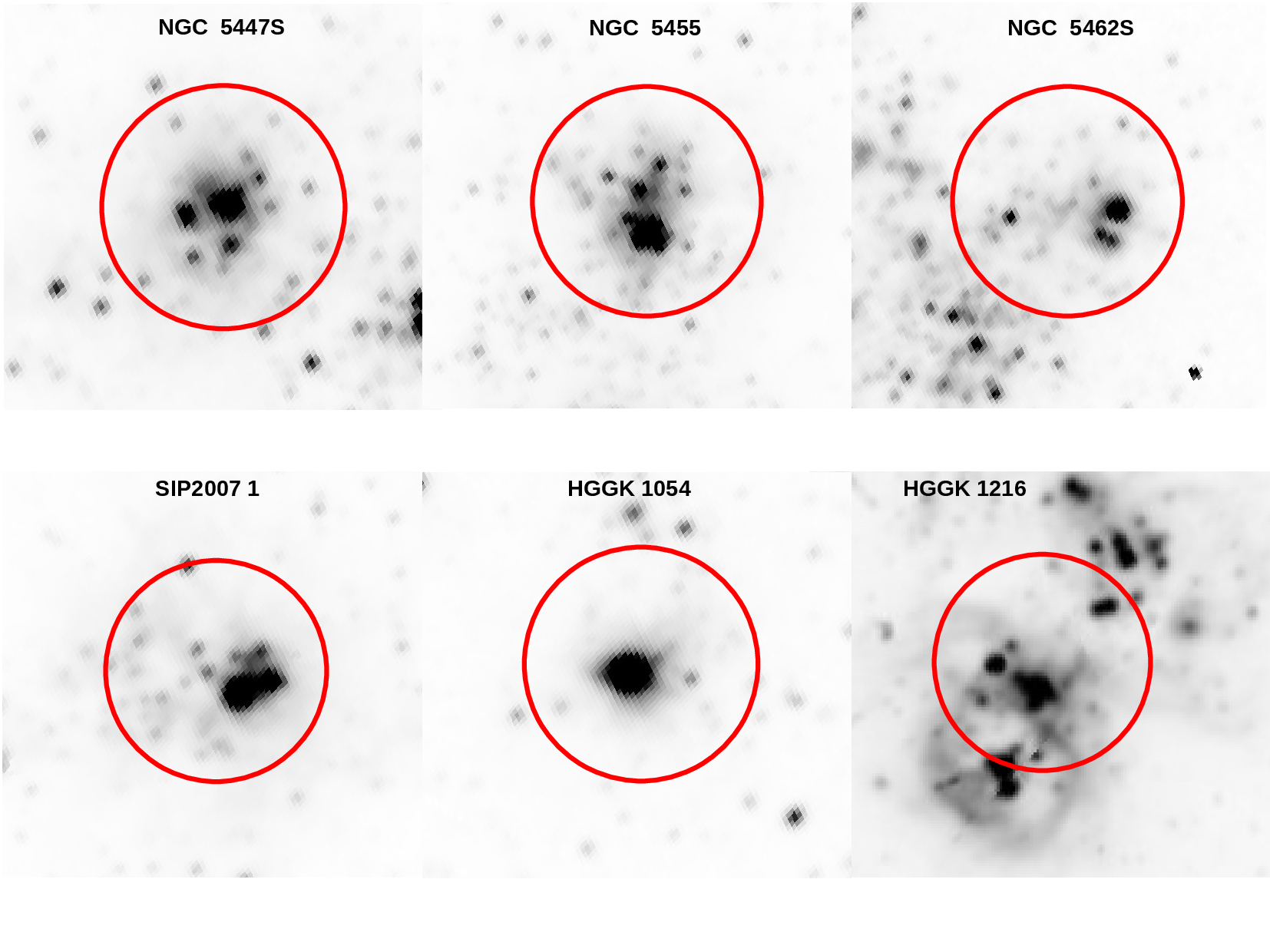}
\caption{Star-forming regions in M101. The red circle shows the position and size of the COS aperture (2.5''). All images are B band from HST, except for HGGK~1216 (V band). North is up, and east is to the left.}
\label{m101}
\end{figure}

\begin{figure}[t]
\centering
\includegraphics[width=0.49\textwidth]{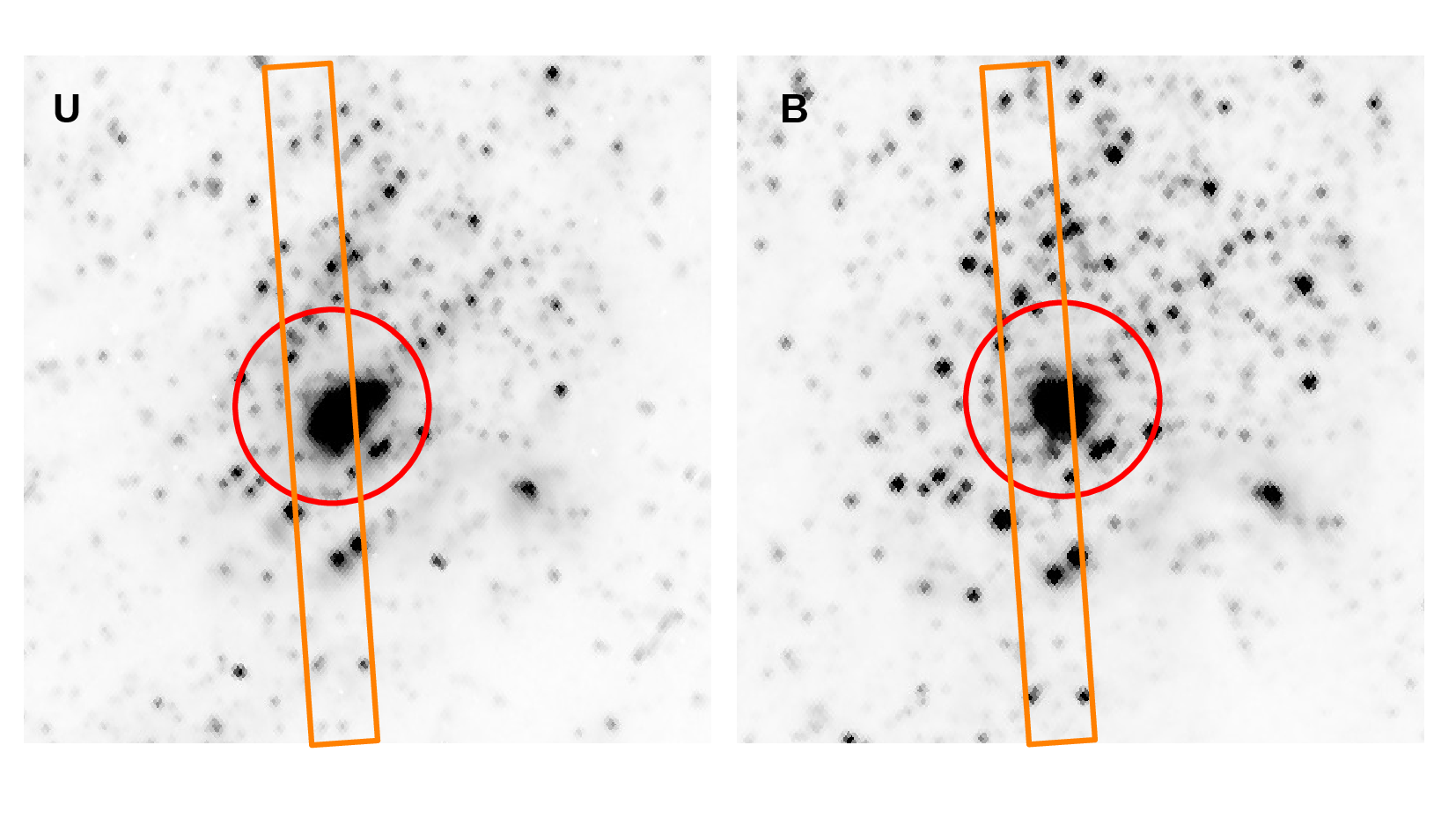}
\includegraphics[width=0.49\textwidth]{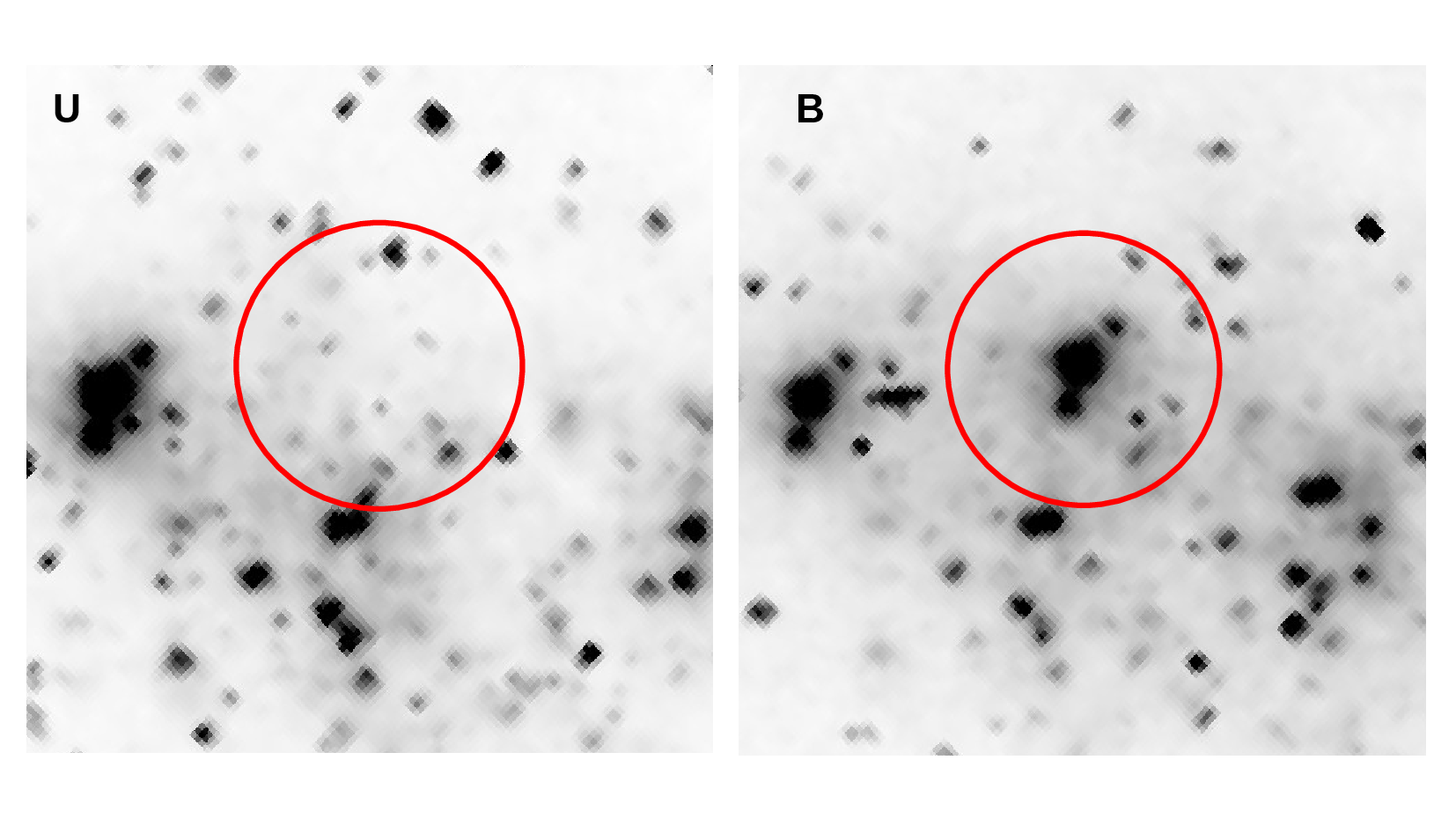}
\caption{Observed fields in NGC~4214 (top) and NGC~4670 (bottom). The red circle is the COS aperture, and the orange rectangle show the STIS aperture. For the latter, the full silt length is not shown (it is out of the frame). The left (right) panels show U-band (B-band) images. North is up, and east is to the left.}
\label{ngc4214_4670}
\end{figure}

We show in Figs.~\ref{m101} and \ref{ngc4214_4670} optical images of the sources in M101, NGC~4214, and NGC~4670. They illustrate the morphology of the sources (similar images for the other sample sources are found in \citet{sidoli06,hc06,smith16,senchyna17,senchyna20,berg22}).

\section{Equivalent widths}
\label{ap_ew}

The wavelength ranges for the EW calculations are given in Table~\ref{tab_ew1}. We collect the EW measurements in Table~\ref{tab_ew2}. The EW of \ion{N}{iii} 4640 + \ion{C}{iii} 4650 is obtained from the difference between EW(blue bump) and EW(\heiiopt). To estimate the EW of \hb\ associated to models, we ran CLOUDY simulations with an ionizing parameter $log U = -2.0$ and an electron density of 200 cm$^{-3}$. The input SED was that of our population synthesis models, and we assumed an LMC-type chemical composition for the gas. Our measurements of nebular EW(\hb) usually agree with those of \citet{senchyna20} for the sources in common. For EW(\heiiuv), we find lower values on average. We used the same spectra and wavelength range as Senchyna et al. for EW(\heiiuv) measurements, and therefore, we attribute the differences to the choice of the continuum. As stated in Sect.~\ref{s_ew}, this critically affects the EW values.

\begin{table}
\caption{Wavelength range for the EW calculations.}
\label{tab_ew1}
\center
\begin{tabular}{l|c}
\hline 
Line ID & Wavelength range \\
        & [\AA] \\
\hline
\heiiuv\ & 1630-1650 \\
\heiiopt\ & 4670-4710 \\
Blue bump & 4615-4710 \\
\hb    &  4850-4875 \\
Red bump & 5780-5850 \\
\hline 
\end{tabular}
\end{table}

\begin{table*}[ht]
\begin{center}
\caption{Equivalent widths of models that include VMS, and sources.} \label{tab_ew2}
\begin{tabular}{lrrrrrr}
  \hline
  Model/Source  & EW(blue bump) & EW(\heiiopt) & EW(red bump) & EW(\heiiuv) & EW(H$\beta^{nebular}$) \\
                &  [\AA]  &  [\AA]  &  [\AA]  &  [\AA] &   [\AA] \\
\hline
burst 0 Myr    &    2.12  &  1.70   &  0.75   &  0.04  & 698 \\
burst 0.5 Myr  &    3.51  &  2.90   &  2.25   &  0.59  & 681 \\
burst 1 Myr    &    7.20  &  6.16   &  5.87   &  1.55  & 661 \\
burst 1.5 Myr  &   15.18  &  13.78  &  1.45   &  2.73  & 613 \\
burst 2 Myr    &   24.79  &  21.78  &  1.62   &  3.69  & 536 \\
burst 2.5 Myr  &   40.02  &  31.29  &  1.44   &  7.57  & 517 \\
CSF 3 Myr      &   17.95  &  14.71  &  3.05   &  3.07  & 560 \\
CSF 5 Myr      &   11.28  &   9.23  &  3.93   &  1.79  & 419 \\
CSF 8 Myr      &    8.28  &   6.67  &  2.52   &  0.72  & 337 \\
CSF 10 Myr     &    7.93  &  6.16   &  2.15   &  1.06  & 301\\
CSF 20 Myr     &    4.81  &  3.90   &  0.66   &  0.39  & 220 \\
CSF 50 Myr     &    3.03  &  2.51   &  0.20   &  0.78  & 145 \\
\hline
R136            &  13.56  &  12.48  &  1.55   & 3.81  & -- \\
NGC~3125-A1     &  18.71  &  12.89  &  1.27   & 5.54  & 108  \\
NGC~5253-5      &  7.89   &  4.66   &  3.05   & 1.58  & 322  \\
Tol~89-A        &  7.68   &  4.10   &  6.30   & 3.89  & 43  \\
NGC~5447S       &  3.07   &  4.07   &  --     & 1.66  & 153  \\
NGC~5455        &  8.78   &  5.32   &  3.47   & 2.35  & 231  \\
NGC~5462S       &  -0.18  & -0.11   &  3.43   & 0.84  & 278  \\
SIP2007-1       &  12.26  &  6.24   &  0.10   & 1.77  & 139  \\
HGGK~1054       &  10.41  &  5.88   &  1.34   & 2.04  & 191  \\
HGGK~1216       &  2.50   &  1.76   &  0.66   & 0.88  & 120  \\
NGC~4214        &  1.43   &  0.87   &  --     & 4.51  & --  \\
NGC~4670        &  --     &  --     &  --     &  0.24 & --  \\
Mrk~33           &  1.98          &  1.19  &    0.89  & -0.20 & 32    \\
J0036$-$3333    &  0.44   &  0.69  &    2.04  & -0.21 & --   \\ 
J0115-sb49      &  7.38   &  5.16  &    7.00  &  1.64 & 225  \\ 
J0823$+$2806    &  4.32   &  2.69  &    3.24  &  1.27 & 82  \\          
J0942-sb80      &  8.30   &  4.72  &    5.59  &  1.17 & 248  \\ 
J1105$+$4444    &  3.31   &  2.14  &    2.73  &  1.99 & 103  \\ 
J1129-sb179     &  7.47   &  4.85  &    2.68  &  3.68 & 202  \\ 
J1132-sb125     &  5.34   &  2.77  &    1.04  &  0.78 & 149   \\                
J1157$+$3220    &  3.80   &  1.89  &    3.17  &  1.64 & 49  \\  
J1200$+$1343    &  9.78   &  6.42  &    1.87  &  3.32 & 223  \\ 
J1215-sb191     &  13.46  &  9.32  &    1.84  &  4.71 & 432   \\          
J1304-sb9       &  13.46  &  6.76  &    12.12 &  1.96 & 171 \\  
J1314-sb153     &  9.43   &  6.52  &    4.73  &  1.95 & 304 \\  
J1428$+$1653    &  1.16   &  0.31  &    2.07  &  2.19 & 43     \\               
J1525$+$0757    &  1.35   &  0.63  &    1.87  &  3.85 & 26   \\         
\hline
\end{tabular}
\tablefoot{Positive values correspond to emission, and negative values correspond to absorption. The wavelength ranges over which EWs are calculated are given in Table~\ref{tab_ew1}. The first 12 lines correspond to models, and the following 15 lines correspond to sample sources. The last column is the EW of the nebular \hb\ line in the observed spectra or the CLOUDY models. See the text for a discussion.}
\end{center}
\end{table*}

\begin{figure*}[]
\centering
\includegraphics[width=0.49\textwidth]{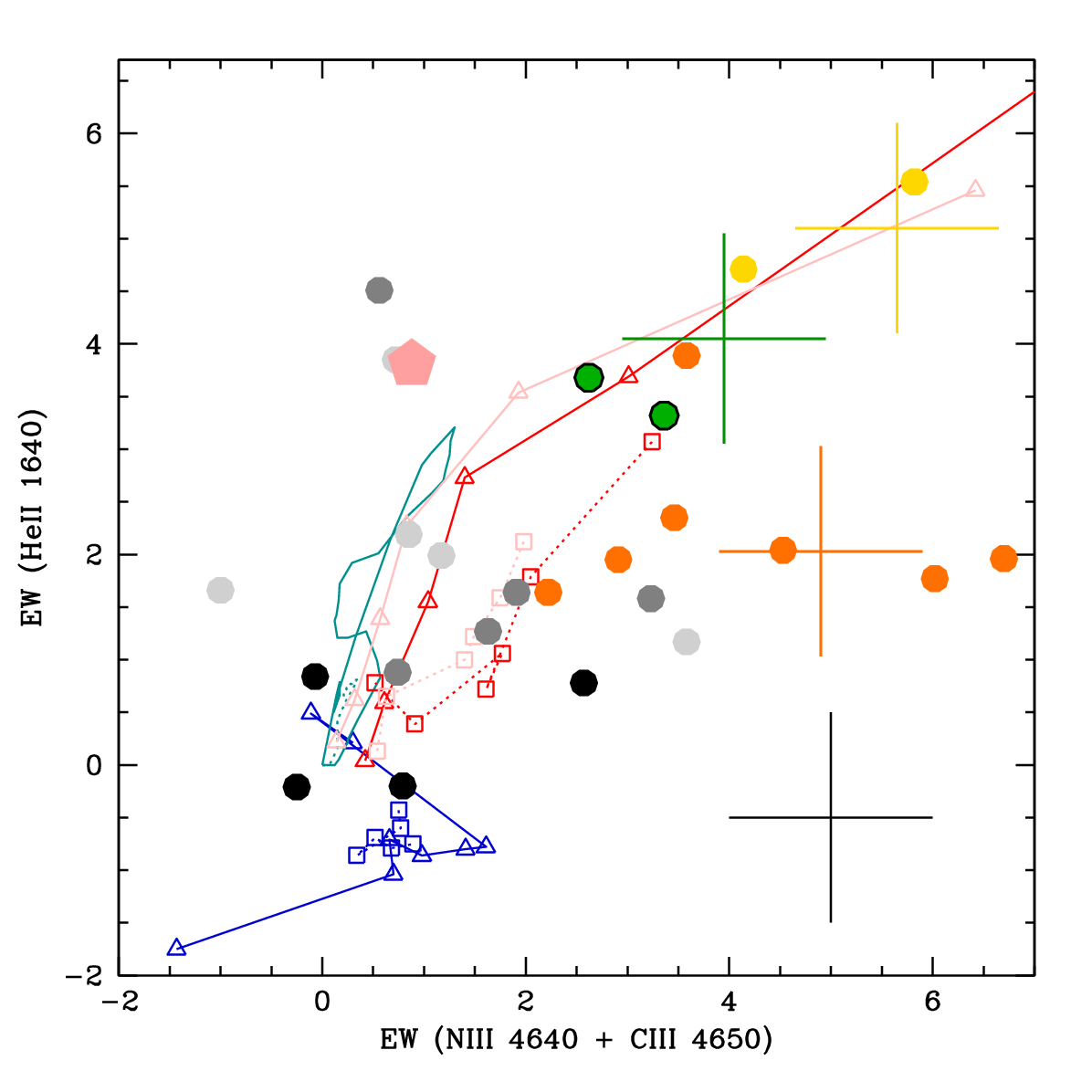}

\includegraphics[width=0.49\textwidth]{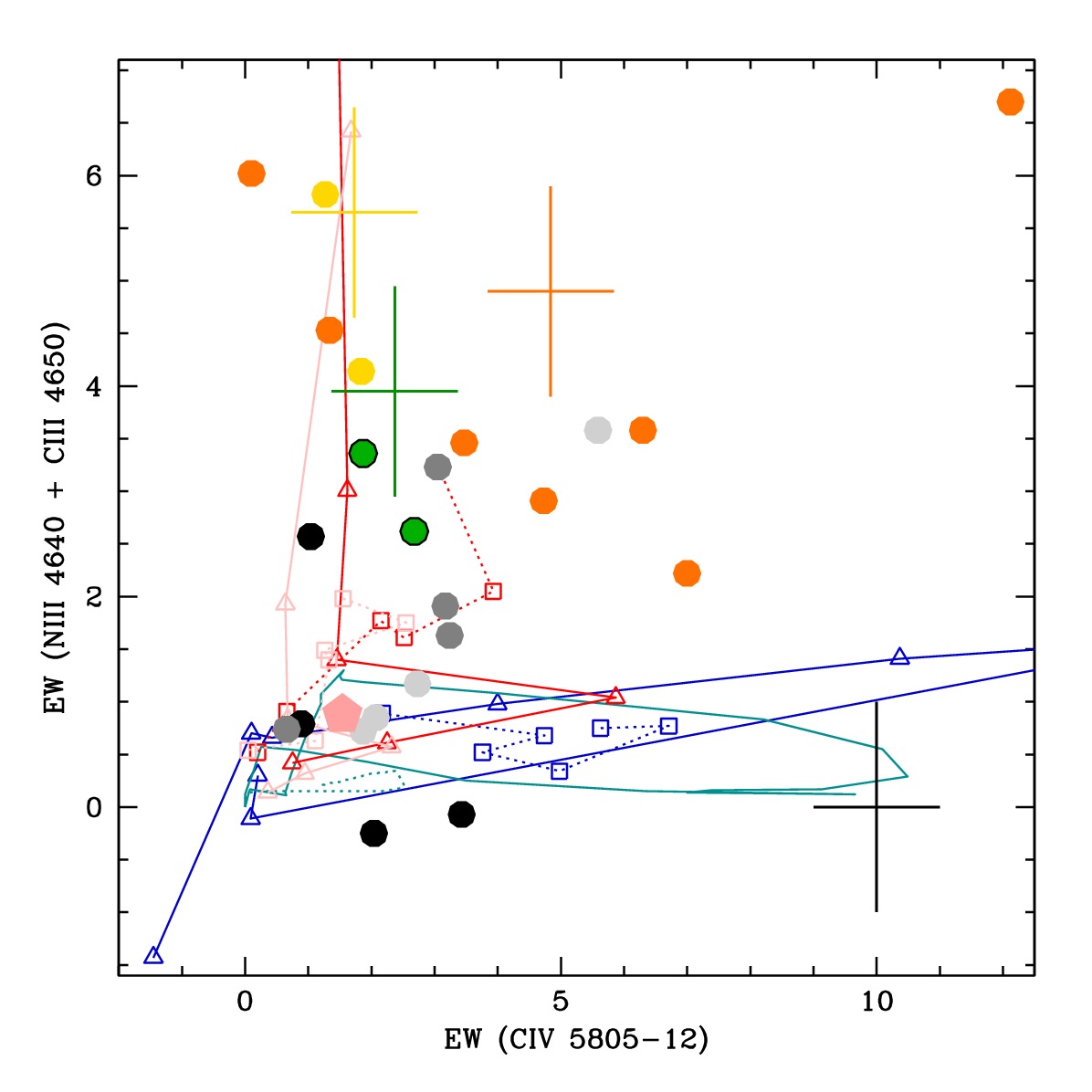}
\includegraphics[width=0.49\textwidth]{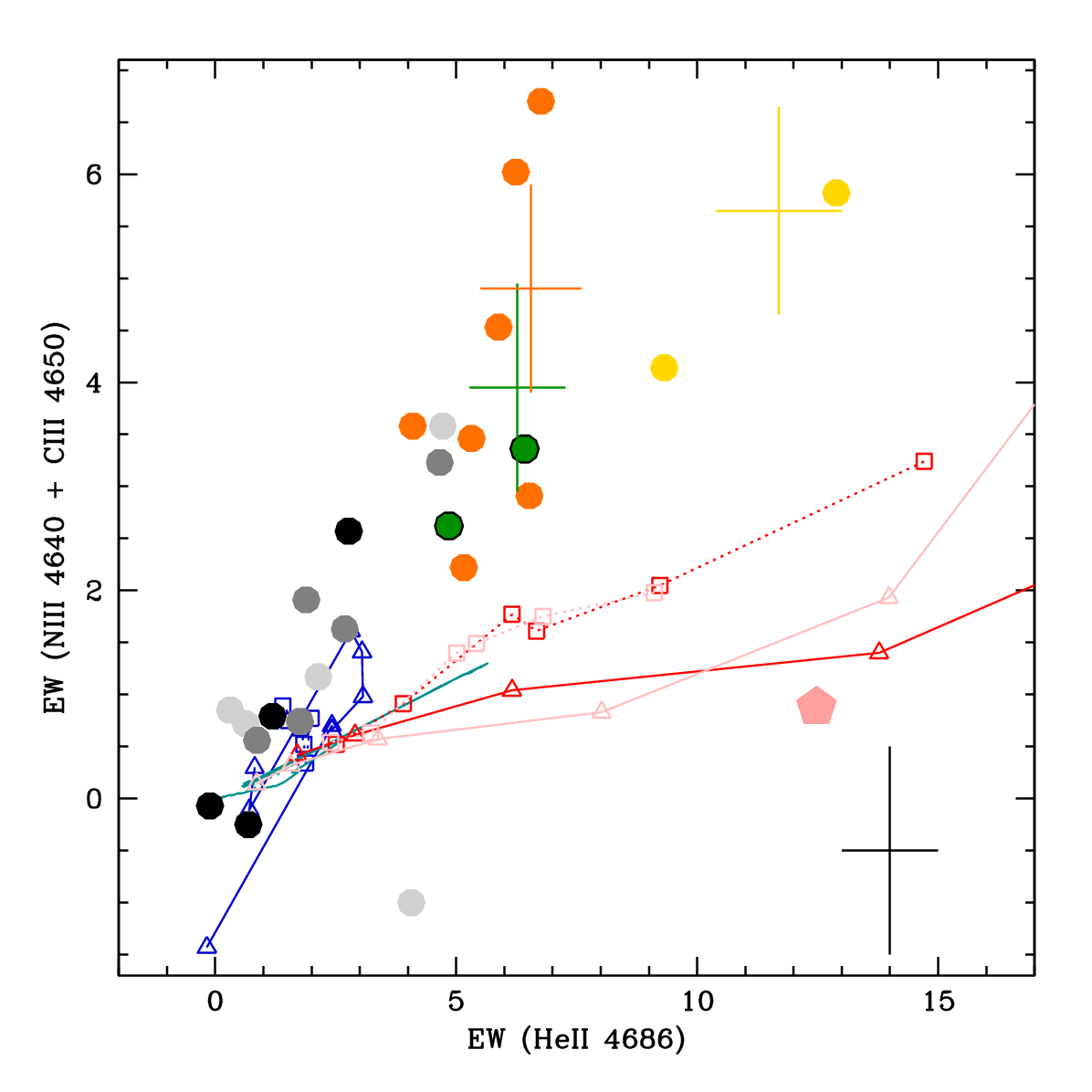}
\caption{Same as Fig.~\ref{fig_ew}, now including models. The solid lines and open triangles are models for star formation bursts. The dotted lines and open squares are models for continuous star formation. The red models include VMS (present study), the blue models are BPASS, and the cyan models are from \citet{sv98}. The pink models are the VMS models that include nebular continuum emission. }
\label{fig_ew2}
\end{figure*}

Fig.~\ref{fig_ew2} shows the comparison between models and observations for the EW of various lines. The models are discussed in Sect.~\ref{s_popshort}.

\section{Classification}
\label{ap_class}

\begin{figure*}[t]
\centering
\includegraphics[width=0.99\textwidth]{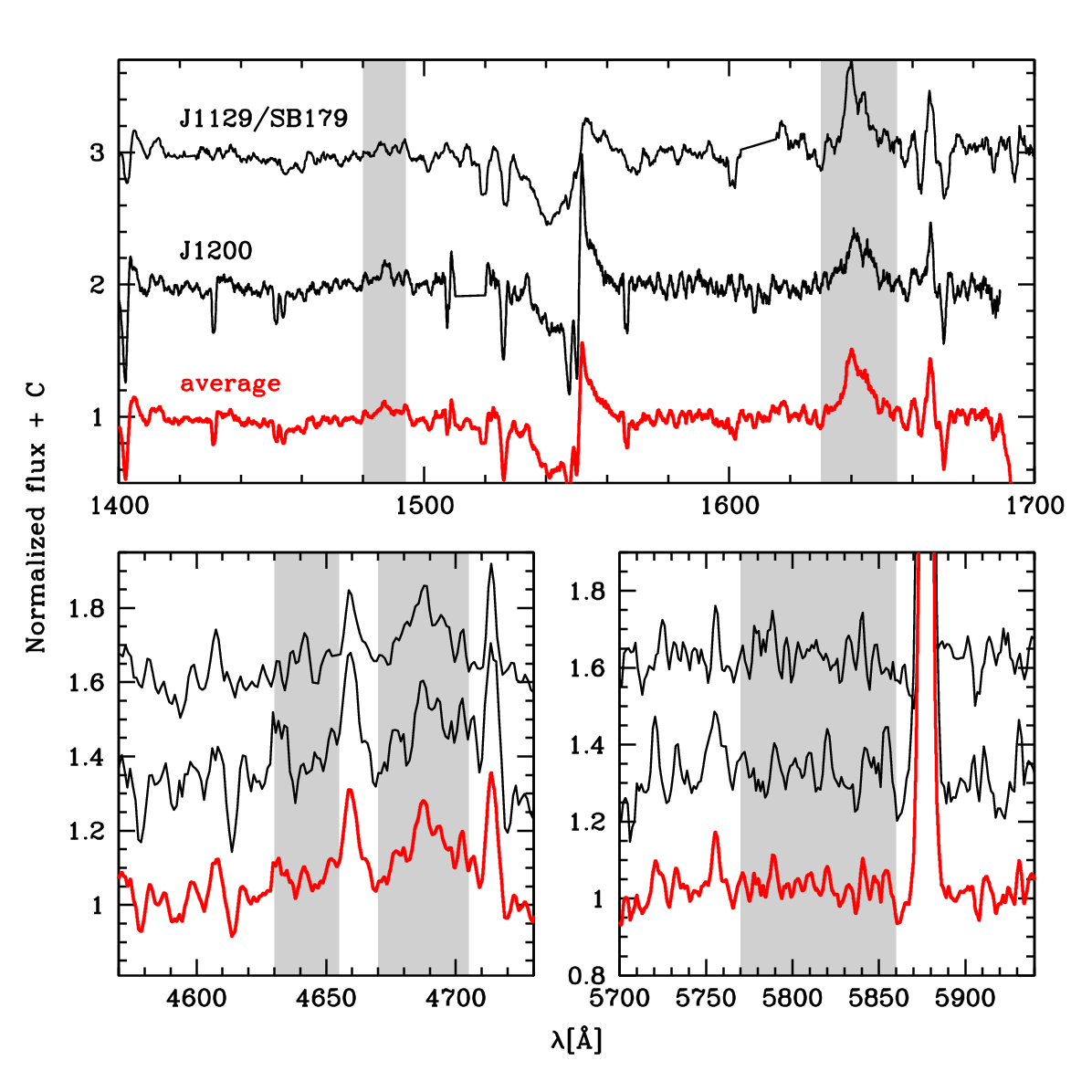}
\caption{Spectra of VMS sources, with the average spectrum in red. The template VMS source R136 is not included in the figure, nor is the average shown. The main stellar lines are highlighted in gray.} 
\label{fig_vms}
\end{figure*}

\begin{figure*}[t]
\centering
\includegraphics[width=0.99\textwidth]{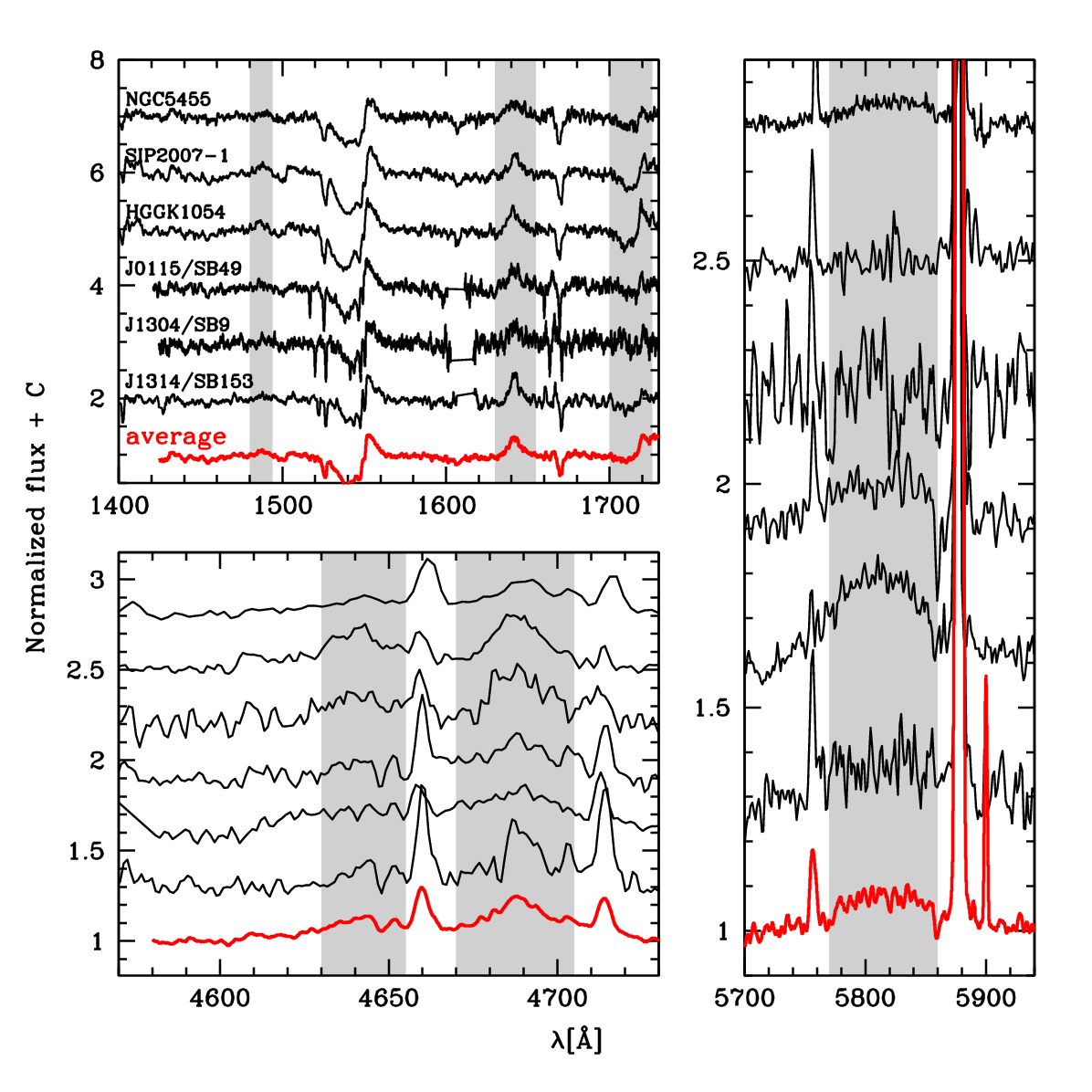}
\caption{Spectra of WR sources, with the average spectrum in red. The template WR source Tol~89-A is not included in the figure, nor is the average shown. The main stellar lines are highlighted in gray.}
\label{fig_wr}
\end{figure*}

\begin{figure*}[t]
\centering
\includegraphics[width=0.99\textwidth]{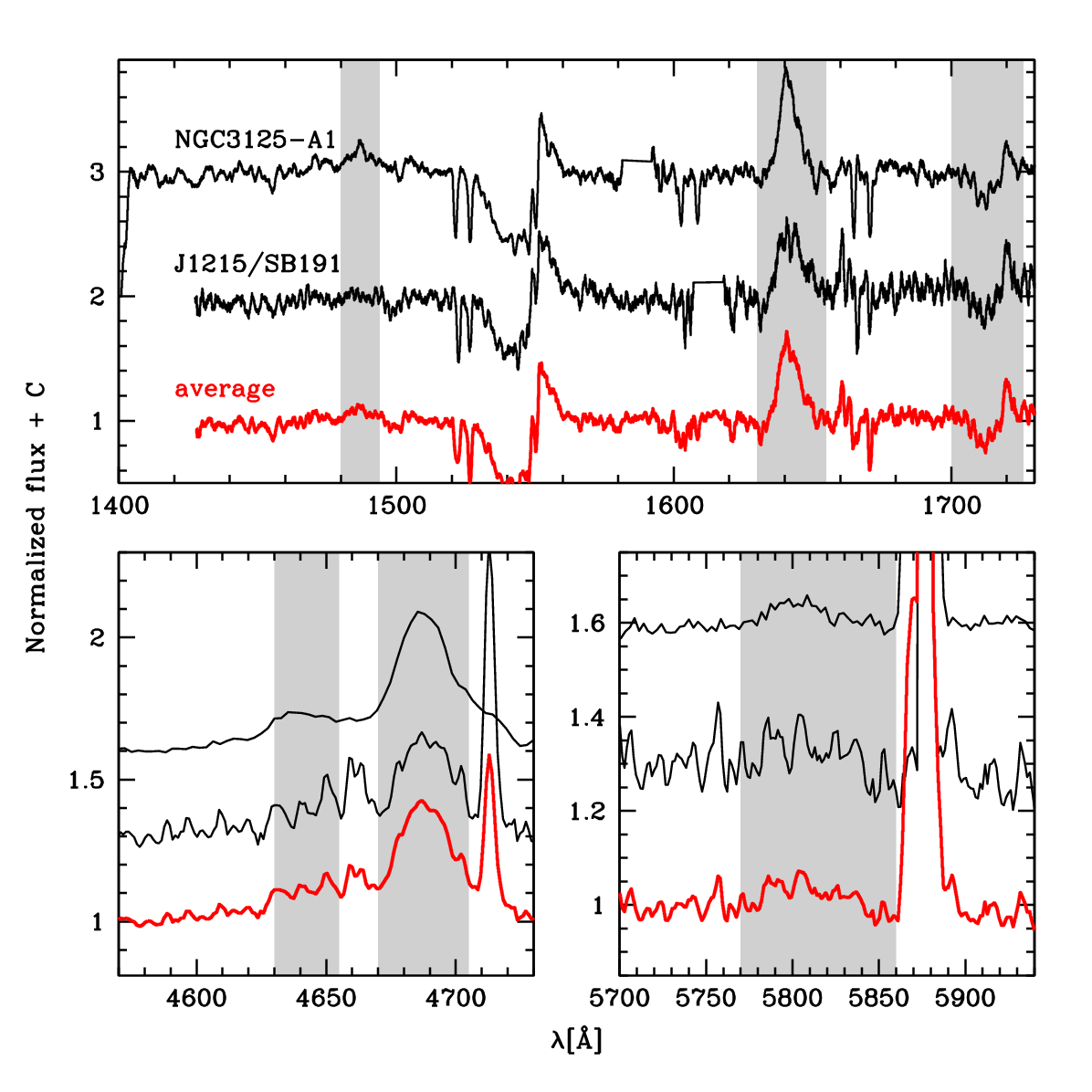}
\caption{Spectra of VMS or WR sources, with the average spectrum in red. The main stellar lines are highlighted in gray.}
\label{fig_vmsORwr}
\end{figure*}

\begin{figure*}[t]
\centering
\includegraphics[width=0.99\textwidth]{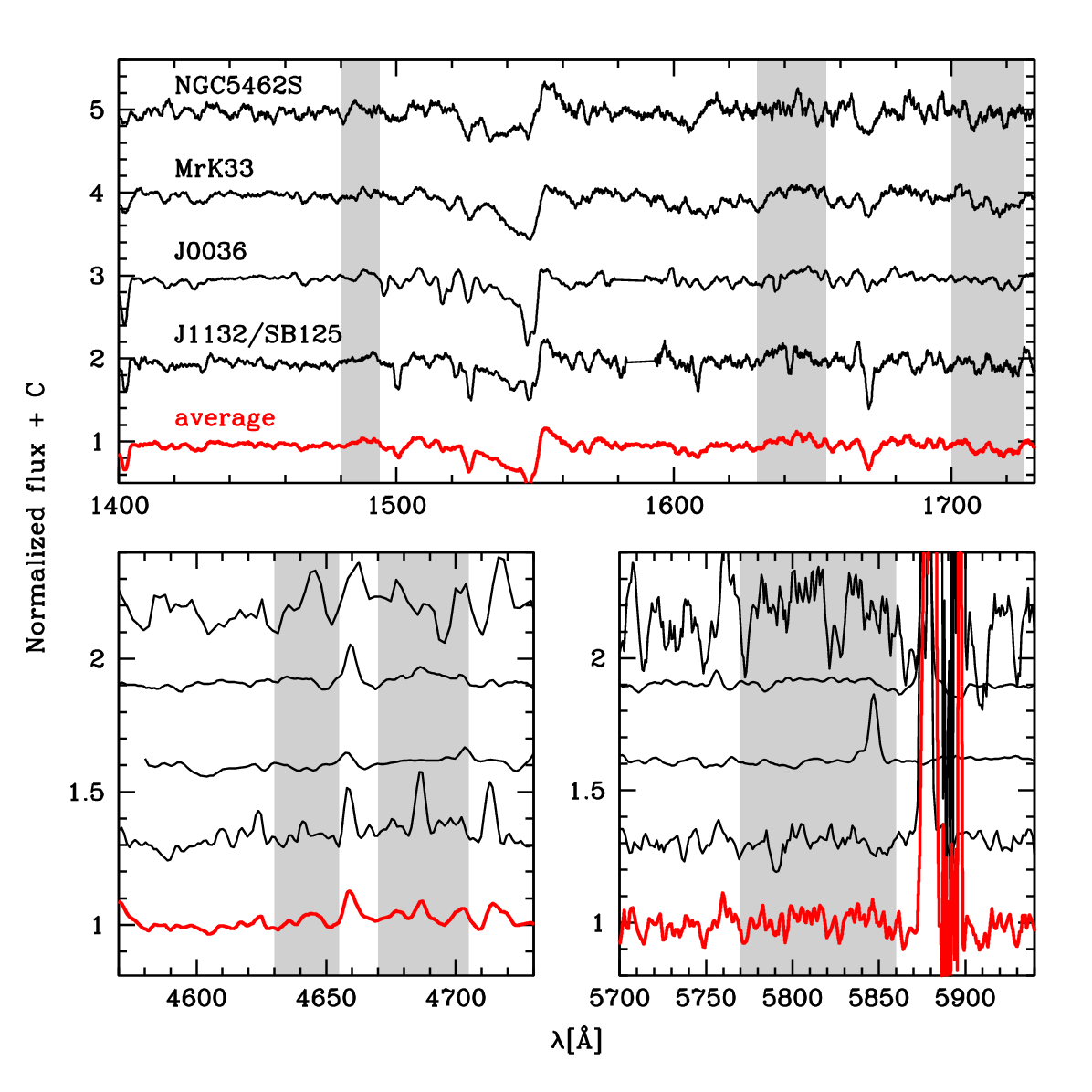}
\caption{Spectra of sources with neither VMS nor WR signatures, with the average spectrum in red. The main stellar lines are highlighted in gray.}
\label{fig_none}
\end{figure*}

\begin{figure*}[t]
\centering
\includegraphics[width=0.99\textwidth]{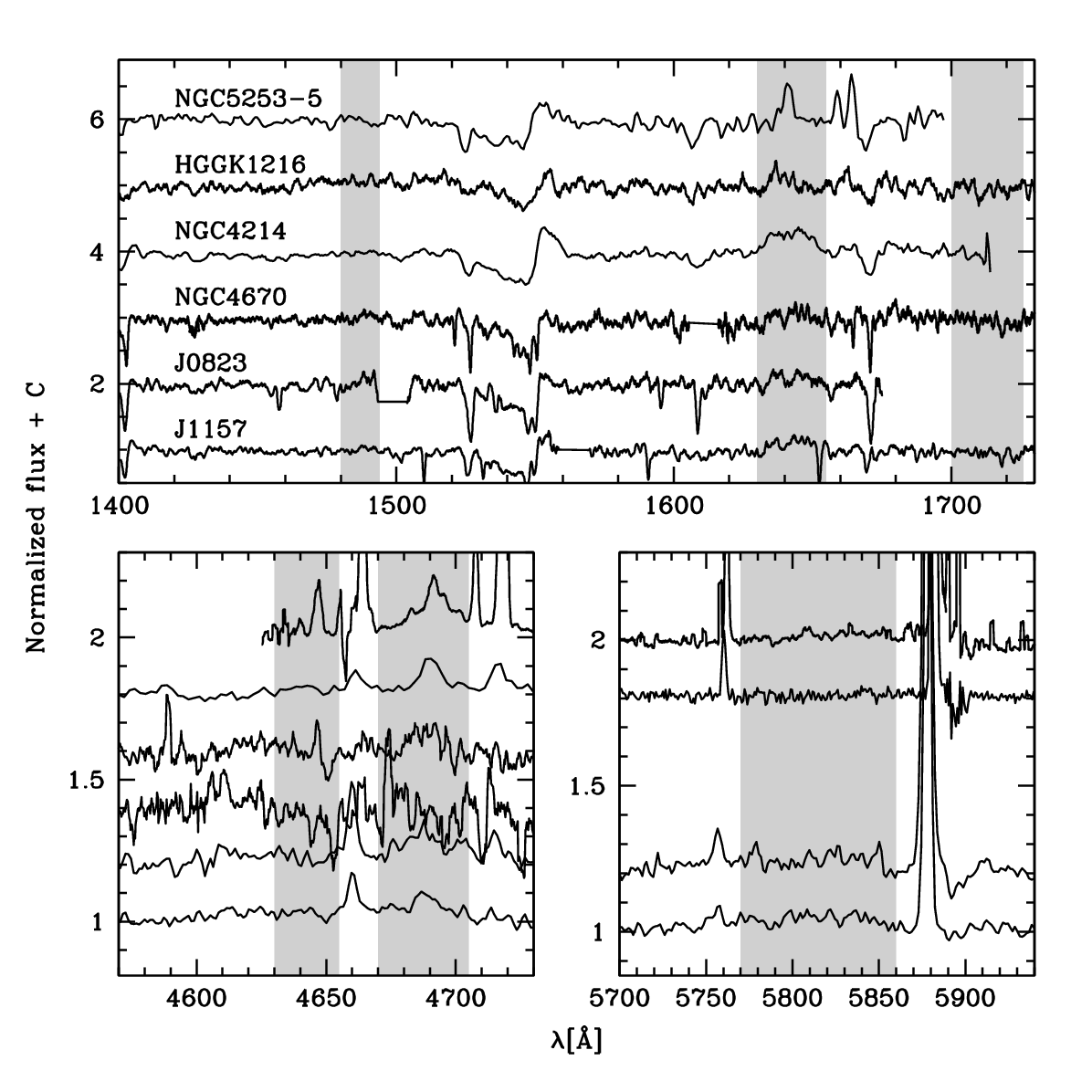}
\caption{Spectra of sources with unknown classification. The main stellar lines are highlighted in gray.}
\label{fig_unk}
\end{figure*}

\begin{figure*}[t]
\centering
\includegraphics[width=0.99\textwidth]{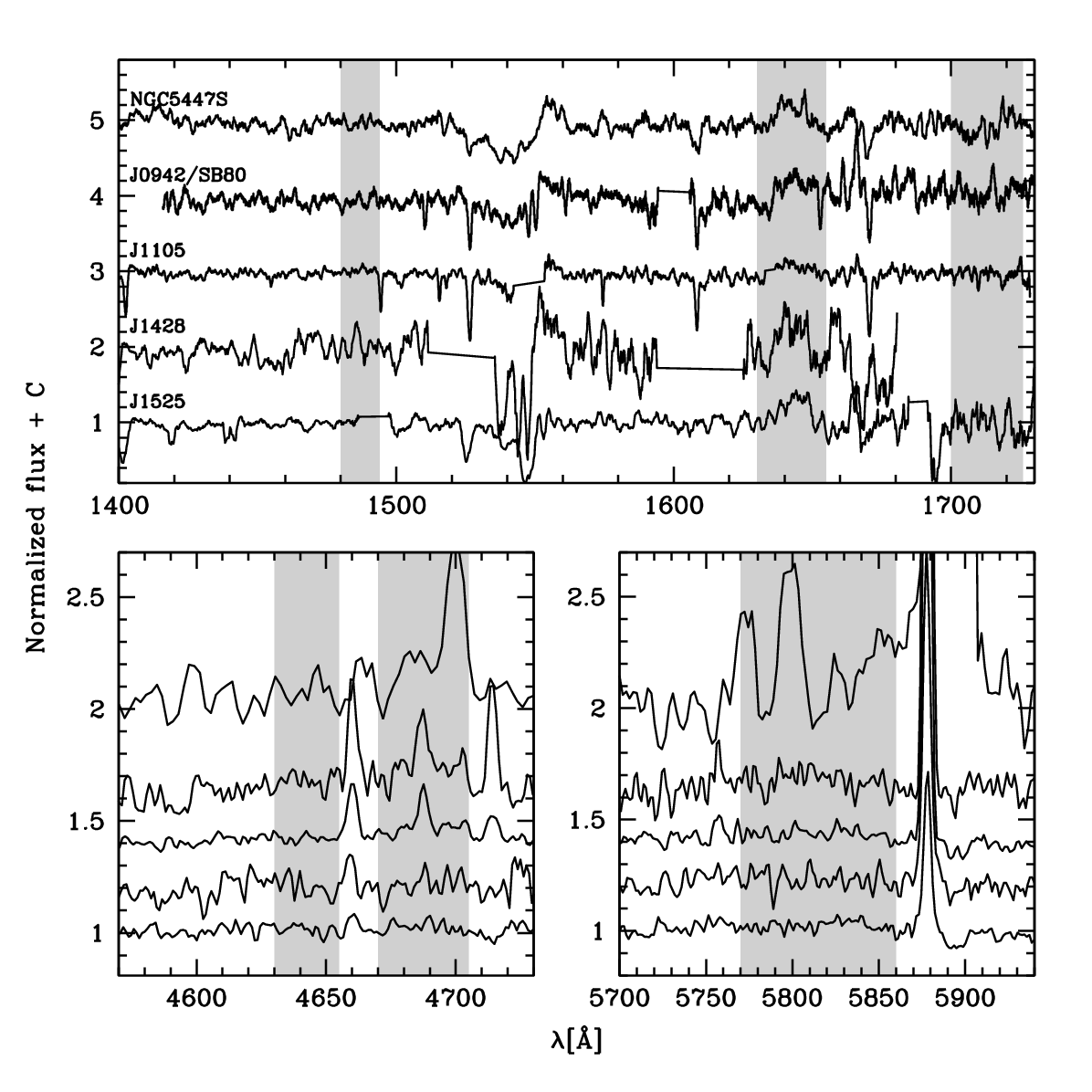}
\caption{Spectra of sources for which the data do not lead to a classification, that is, the sources with a questionmark in Table~\ref{tab_classif}. The main stellar lines are highlighted in gray.}
\label{fig_qmark}
\end{figure*}

Figs.~\ref{fig_vms} to \ref{fig_qmark} shows the UV and optical spectra (blue and red bumps) of our sample sources. We grouped them according to the classification results, as described in Sect.~\ref{s_emp}.

\end{appendix}

\end{document}